\documentclass[a4paper,11pt]{article}
\pdfoutput=1 

\usepackage{jinstpub} 

\usepackage{lineno}

\usepackage{booktabs, multirow} 
\usepackage{soul}
\usepackage[table]{xcolor} 
\usepackage{changepage,threeparttable} 
\usepackage{siunitx}
\usepackage{subcaption}
\usepackage{graphicx}
\usepackage{dcolumn}
\usepackage{bm}

\title{
Measurement of Sudden Beam Loss Events Using Bunch-by-Bunch BPMs at SuperKEKB
}

\author[1, 2]{R. Nomaru,}
\author[2, 3]{G. Mitsuka,}
\author[4]{and L. Ruckman,}
\affiliation[1]{
    Department of Physics, The University of Tokyo, Bunkyo, Tokyo 113-0033, Japan
}
\affiliation[2]{
    KEK, Oho, Tsukuba, Ibaraki 305-0801, Japan
}
\affiliation[3]{
    SOKENDAI, Shonan Village, Hayama, Kanagawa 240-0193, Japan
}
\affiliation[4]{
    SLAC National Accelerator Laboratory,
    2575 Sand Hill Road, M/S 96,
    Menlo Park, CA 94025, U.S.A.
}

\emailAdd{nomaru@post.kek.jp}

\abstract{

\noindent

At SuperKEKB, Sudden Beam Loss (SBL) events pose a significant challenge to stable accelerator operation.
To investigate and better understand SBL, we have developed a new Bunch Oscillation Recorder (BOR).
Using this newly developed BOR, we successfully observed SBL events and conducted a detailed analysis of the recorded data.
By analyzing the patterns of bunch position oscillations and charge loss, we found a strong correlation between SBL events and pressure burst phenomena occurring inside the vacuum chamber.
These pressure bursts are known to accompany almost all SBL events, and our analysis shows that the bunch position oscillation patterns vary depending on the location of the pressure burst.
Our observations suggest that bunch positions begin oscillation under some influence at the location of the pressure burst.
These observations and analyses have significantly advanced our understanding of the causes and mechanisms behind SBL.
}

\begin{document}
\maketitle
\flushbottom
\setlength{\parskip}{0pt}


\section{Introduction}
The SuperKEKB accelerator~\cite{SuperKEKBTDR} collides 4~GeV positrons with 7~GeV electrons at extremely high luminosity, producing large numbers of B~mesons, D~mesons, $\tau$~leptons, etc. for the Belle~II experiment~\cite{10.1093/ptep/ptz106}.
In December 2024, SuperKEKB achieved a world-record instantaneous luminosity of $5.1\times10^{34}~\mathrm{cm}^{-2}\mathrm{s}^{-1}$.
The long-term goal is to exceed this record by more than an order of magnitude.
However, a phenomenon known as Sudden Beam Loss (SBL) has emerged as a major obstacle to stable accelerator operation~\cite{ikeda:ipac2025-mocd3}.

SBL refers to the abrupt loss of a large portion of the stored beam within tens of microseconds, ultimately resulting in a beam abort.
Since the response time of the abort system is typically 10–20~µs (about 1–2 turns), the beam cannot be aborted quickly enough, and significant beam loss strikes accelerator components before the abort is completed.
Such uncontrolled losses have caused serious damage, including to the innermost silicon detectors of Belle~II and collimator heads, as well as quenching of the superconducting final-focus magnets located near the collision point.
Understanding the mechanism of SBL is therefore essential for both machine protection and future luminosity upgrades.

So far, two main diagnostics have been used to study SBL: loss monitors, which detect radiation from lost particles, and vacuum gauges, which measure pressure bursts~\cite{Terui:2018pae} inside the beam chamber.
Pressure bursts are observed in most SBL events and are thought to occur at locations suspected to be the source points of SBL~\cite{ikeda:ipac2025-mocd3}.
While indispensable, these devices cannot capture the detailed dynamics of individual bunches leading up to an SBL event.
Loss monitors integrate over many bunches and locations, and vacuum gauges have response times too slow to resolve the microsecond-scale evolution of SBL.
As a result, the underlying processes—such as the onset of bunch oscillations or the sequence of charge depletion—remain not well understood.

To overcome these limitations, we have developed a new Bunch Oscillation Recorder (BOR) using AMD/Xilinx Radio Frequency System-on-Chip (RFSoC) technology~\cite{Nomaru:2024qls}.
The BOR is a dedicated diagnostic that records bunch-by-bunch beam position and charge for all stored bunches during the final $\sim$100 turns before a beam abort.
This enables direct observation of individual bunch dynamics during SBL events, which has not been possible with conventional diagnostics.
Although multiple BOR types are installed at SuperKEKB, in this paper we refer specifically to the newly developed RFSoC-based BOR, hereafter referred to simply as “BOR.”

The main purpose of this study is to demonstrate the diagnostic capability of the BOR — to verify whether it can serve as a new “eye” for observing fast beam instabilities and to clarify what new information it can reveal that has been inaccessible with conventional BPMs, loss monitors, or vacuum gauges.
Through direct observations of SBL events, we show that the BOR provides complementary and previously unattainable information compared with these existing diagnostics.
With the BOR, we can identify when and where bunch oscillations start, quantify their amplitudes and durations, and determine in which section of the ring the charge loss occurs.
By correlating these results with simultaneous measurements from loss monitors and vacuum gauges, we obtain new insights into the causal chain linking pressure bursts, beam instabilities, and sudden beam losses.
This paper presents the first systematic observations and analyses of SBL events using the BOR at SuperKEKB, establishing its role as a powerful and unique diagnostic tool that deepens our understanding of the temporal evolution of SBL phenomena and provides valuable perspectives for future beam instability studies.

At the Large Hadron Collider (LHC), unidentified falling objects (UFOs)—micrometer-sized dust particles interacting with the circulating beam—had been reported to cause rapid beam losses~\cite{Baer:2011mf, Lindstrom:2020hks, Auchmann:2016upc}.
Although the beam particle species accelerated at SuperKEKB and the LHC are different, it is important to examine whether the underlying mechanism is similar to that of dust-induced losses or fundamentally different.
No previous study has analyzed such fast beam-loss phenomena using bunch-by-bunch position data.
This work therefore not only sheds light on the mechanisms behind SBL at SuperKEKB but also provides insights of potential relevance to other accelerator facilities.

This paper is organized as follows.
Section~\ref{01_Methodology_of_SBL_Observation} describes the experimental setup for SBL observation.
Section~\ref{02_BOR_data_analysis_for_SBL_events} presents analyses of SBL events recorded by the BOR, including charge loss distributions, oscillation onset times, and oscillation amplitudes.
Section~\ref{03_Analysis_Using_Multiple_Monitoring_Systems} combines BOR results with loss monitors and vacuum gauges, highlighting correlations with pressure bursts and bunch oscillations.
Finally, Section~\ref{04_Summary} summarizes the key findings and discusses prospects for further studies.

\section{Methodology of SBL Observation}
\label{01_Methodology_of_SBL_Observation}

This section provides an overview of the SuperKEKB main ring and describes the setup of the BOR used for observing SBL events.

\begin{figure}[htbp]
\centering
\includegraphics[width=0.7\linewidth]{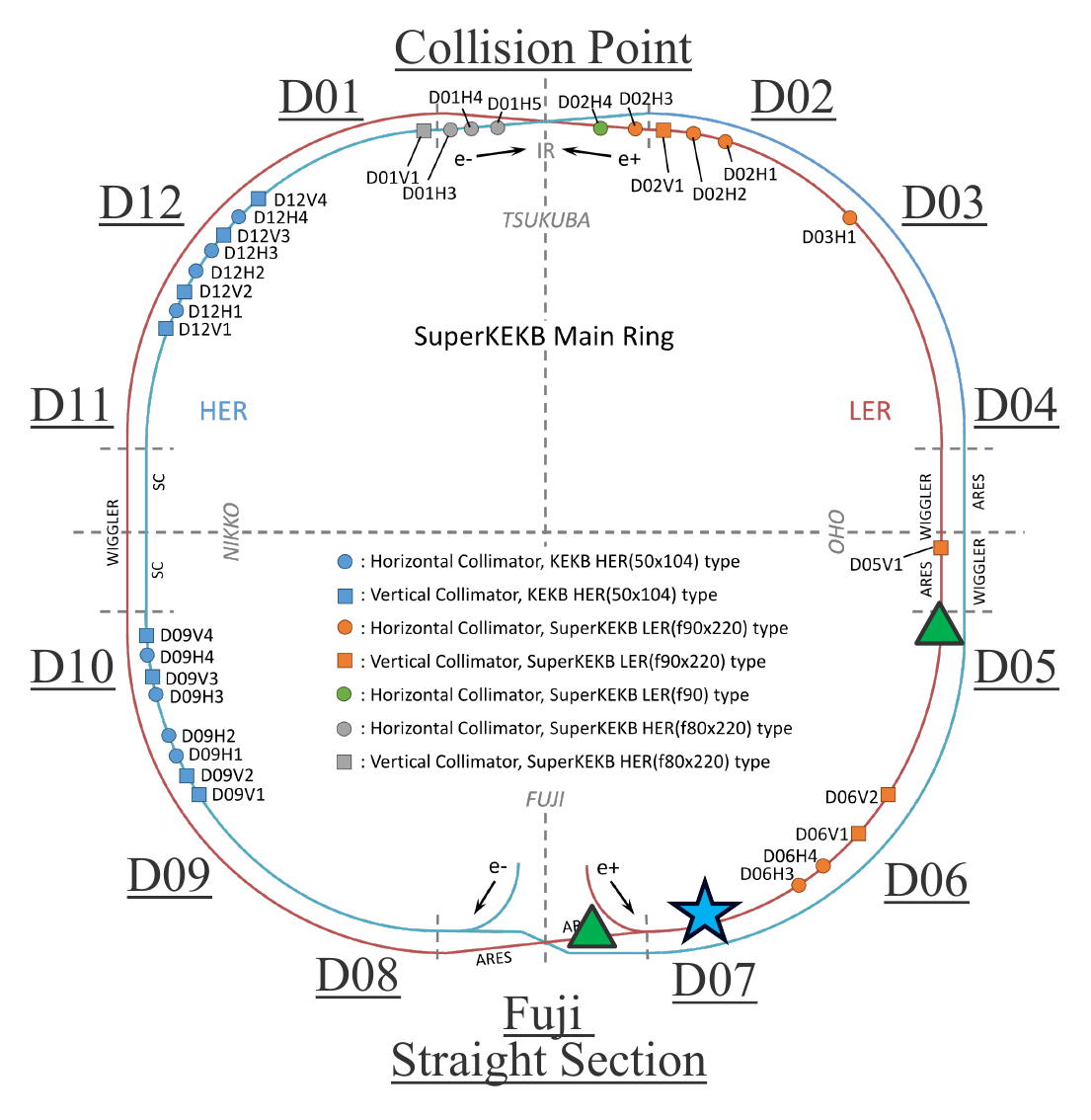}
\caption{\label{fig:collimator_location} Schematic diagram of the SuperKEKB main ring~\cite{Ishibashi:2020rgt}. The red ring represents the LER, and the light blue ring represents the HER. Squares along the ring indicate the locations of vertical collimators, and circles indicate the locations of horizontal collimators. In the LER the blue star marks the beam dumps, and green triangles indicate the locations of the BORs.}
\end{figure}

The SuperKEKB main ring consists of two storage rings: the 4~GeV positron ring (Low Energy Ring: LER) and the 7~GeV electron ring (High Energy Ring: HER).
Typical machine parameters are summarized in Table~\ref{table:kasokuki_param}, and a schematic of the main ring is shown in Fig.~\ref{fig:collimator_location}.
The beam circulates counterclockwise in the LER and clockwise in the HER.
The main ring is divided into 12 sections labeled D01 through D12, proceeding clockwise around the ring (as indicated around the perimeter in Fig.~\ref{fig:collimator_location}).

\begin{table}[hbt]
\centering
\begin{tabular}{lr|cc}
\hline
               & \multicolumn{1}{l|}{}                & LER     & HER     \\ \hline
Beam energy          & $E$ {[}GeV{]}                        & 4.0                         & 7.0                         \\
Circumference              & $C$    {[}m{]}                        &             \multicolumn{2}{c}{3016}                        \\
Harmonic number       & $h$                                   &             \multicolumn{2}{c}{5120}                        \\
RF frequency             & $f_{\rm{RF}}$    {[}MHz{]}           &             \multicolumn{2}{c}{509}                        \\
Revolution frequency             & $f_0$    {[}kHz{]}           &             \multicolumn{2}{c}{100}                        \\
Minimum bunch spacing          &      {[}ns{]}                          & \multicolumn{2}{c}{4}                                  \\
Beam current          & $I$ {[}A{]}                          & 1.632                         & 1.259                         \\
Number of bunches           & $n_b$                                & \multicolumn{2}{c}{2346}                                  \\
Charge per 1 mA bunch &  {[}nC{]}         & \multicolumn{2}{c}{10} \\
Horizontal beta function at IP   & $\beta_x^{\ast}$ {[}mm{]}            & 60                          & 60                       \\
Vertical beta function at IP   & $\beta_y^{\ast}$ {[}mm{]}            & 1.0                           & 1.0                     \\
Horizontal tune & $\nu_x$                              & 44.525                      & 45.531                      \\
Vertical tune & $\nu_y$                              & 46.589                      & 43.599                      \\
Vertical beam size at IP  & $\sigma_y^{\ast}$ {[}nm{]}                  &   \multicolumn{2}{c}{$\sim 300$}     \\
Bunch length         & $\sigma_z$ {[}mm{]}                  &    \multicolumn{2}{c}{$\sim 6$}          \\
Instantaneous Luminosity & $L$  {[}$\rm{cm}^{-2}\rm{s}^{-1}${]} & \multicolumn{2}{c}{5.1$\times10^{34}$} \\ \hline
\end{tabular}
\caption{Main parameters of the SuperKEKB main ring. Parameters below the beam current correspond to the conditions at the time of the highest recorded luminosity on December 27, 2024~\cite{SuperKEKB_status_and_plan}.}
\label{table:kasokuki_param}
\end{table}

\subsection{Collimator}
A total of 31 collimators are installed in the main ring: 11 in the LER and 20 in the HER.
Their locations are indicated by squares and circles in Fig.~\ref{fig:collimator_location}.
Collimators inherited from the KEKB era (light blue symbols) have a single movable jaw on either the top or bottom (left or right)~\cite{Suetsugu:2003gb}, whereas the newer SuperKEKB-type collimators are equipped with movable jaws on both sides~\cite{Ishibashi:2020rgt}.
Collimators are named by combining the section name with a suffix “H” for horizontal or “V” for vertical, followed by an index number.

\subsection{Vacuum gauge}
To monitor vacuum pressure inside the beam chambers, approximately 600 cold cathode gauges (CCGs) are installed along the main ring spaced approximately 10~m apart, with roughly 300 in each ring~\cite{Terui:2024jue}.
During operation, the vacuum level is maintained at $10^{-8}$ to $10^{-7}$~Pa.
Each CCG is labeled with a name such as “D01\_L01,” where the suffix “L” or “H” denotes whether the gauge is installed in the LER or HER, followed by a serial number indicating its position along the beamline.

\subsection{Loss monitor and beam abort system}
\label{lossmonitor_and_beamabortsystem}
Loss monitors are distributed throughout the main ring and employ various types of sensors, including PIN photodiodes, optical fibers, ionization chambers~\cite{ikeda:ibic14-tupd22}, diamond detectors~\cite{Bacher:2021frk}, and CLAWS (sCintillation Light And Waveform Sensors) using plastic scintillators and SiPMs~\cite{claws_proc}.
These sensors detect radiation produced by beam loss and issue a beam abort request upon detection.
The request is transmitted optically to the central control room, where an abort kicker trigger is generated.
This trigger activates the abort kicker magnets~\cite{mimashi:ipac14-mopro023}, bending the beam into the beam dump.
The LER beam dump is located in the D07 section (indicated by a blue star in Fig.~\ref{fig:collimator_location}) and the HER dump is situated in the Fuji straight section.

Typically, around 2300 bunches circulate in the main ring, divided into two bunch trains separated by an approximately 300~ns gap called the abort gap, which allows time for the kicker magnetic field to rise.
Since the beam completes one revolution in 10~µs, two abort gaps pass every 5~µs.
When a loss monitor detects beam loss, the system waits for the next abort gap, then issues a trigger.
The kicker field rises during the gap, and the bunches following the gap are bent into the beam dump.

\subsection{SBL observation setup}
\label{sbl_observetion_setup}
Two BORs were developed and installed in the LER to observe SBL events~\cite{Nomaru:2024qls}.
In this paper, we focus on SBL events occurring in the LER, as previous observations have shown that SBL events occur more frequently in the LER than in the HER~\cite{Yoshihara:2024vme}.
One BOR was installed in the Fuji straight section (bottom of Fig.~\ref{fig:collimator_location}), referred to as “Fuji-RFSoC.”
The second was installed in the D05 section (right of Fig.~\ref{fig:collimator_location}), referred to as “D5-RFSoC.”
These locations are marked with green triangles in Fig.~\ref{fig:collimator_location}.
It should be noted that the number of BORs was limited to two during this study.
This limitation was primarily due to budgetary constraints and operational restrictions, as measurements were permitted only at BPM ports that did not interfere with routine machine operation.
Therefore, the purpose of this work is not to provide a statistical survey of SBLs over the entire ring, but rather to demonstrate how much new and detailed information can be obtained even from a small number of observation points.

Between the two BORs lies the beam dump (blue star) and the D06 collimator section, which houses two horizontal collimators (D06H3 and D06H4) and two vertical collimators (D06V1 and D06V2).
Upstream of Fuji-RFSoC, there is the D10 section, where wiggler magnets are installed.
The D10 section contains multiple CCGs spaced about 10~m apart, named sequentially from upstream (e.g., D10\_L01, D10\_L02, ...).
Because the LER beam travels counterclockwise, the beamline proceeds from the D10 section through the Fuji-RFSoC, the beam dump, the D06 collimator section, the D5-RFSoC, and finally to the collision point.
Table~\ref{table:beta_nu} summarizes the optics parameters at relevant locations, including CCGs in the D10 section, BOR locations, and collimators in the D06 section~\cite{Ohnishi:eeFACT25}.

\begin{table}[hbt]
\centering
\begin{tabular}{c|cccccc}
\hline
Location         & $s$ (m)   & $\beta_x$ (m) & $\nu_x$ (rad/$\pi$) & $\eta_x$ (mm) & $\beta_y$ (m) & $\nu_y$ (rad/$\pi$) \\ \hline
D10\_L02 CCG   & 773.19  & 29.80         & 22.98    & 5.15          & 6.82          & 23.59    \\
D10\_L03 CCG   & 782.61  & 6.02          & 23.21   & -93.09        & 28.63         & 23.83   \\
D10\_L05 CCG   & 801.46  & 6.07          & 23.72    & -103.45       & 29.59         & 24.30    \\
D10\_L06 CCG   & 810.89  & 29.80         & 23.97    & -8.45         & 6.87          & 24.51    \\
D10\_L07 CCG   & 820.32  & 7.04          & 24.19    & 91.68         & 28.05         & 24.75    \\
D10\_L08 CCG   & 829.74  & 31.22         & 24.42    & 310.7         & 5.38          & 25.01    \\
Fuji-RFSoC   & 1519.06 & 19.13         & 44.45   & 0             & 19.15         & 46.31   \\
D06H3 collimator & 1788.02 & 24.25         & 52.45    & 695.43        & 5.56          & 54.81    \\
D06H4 collimator & 1816.89 & 24.25         & 53.42    & 695.86        & 5.58          & 55.66    \\
D06V1 collimator & 1870.27 & 14.64         & 54.95    & 515.99        & 67.35         & 57.73   \\
D06V2 collimator & 1989.58 & 99.96         & 58.34   & 446.66        & 20.57         & 61.01    \\
D5-RFSoC   & 2161.12 & 7.05          & 63.94    & 0             & 77.18         & 66.69   \\ \hline
\end{tabular}
\caption{Accelerator parameters at the locations mentioned in this paper within the D10 section, at the positions of the Fuji-RFSoC and D5-RFSoC, and at the collimators in the D06 section. Here, $s$ is the distance from the collision point; $\beta_x$ and $\beta_y$ are the horizontal and vertical beta functions; $\nu_x$ and $\nu_y$ are the horizontal and vertical betatron phases with the collision point defined as 0; and $\eta_x$ is the horizontal dispersion function.}
\label{table:beta_nu}
\end{table}

In this paper, a beam abort is classified as an SBL event if the maximum charge loss among the bunches exceeds 5\%.
The Fuji-RFSoC recorded 117 SBL events between October 13 and November 28, 2024.
The D5-RFSoC recorded 58 events between October 26 and November 28, 2024.
During this period, SuperKEKB operated in two optics configurations.
From October 13 to 23, a “Detuned Optics” setting was used, with relaxed beta functions at the collision point ($\beta_y^* = 48.6$~mm, $\beta_x^* = 384$~mm), and no collisions were performed.
After October 24, the optics were switched to enable collisions, with the vertical beta function squeezed to $\beta_y^* = 1$~mm.
The collimator apertures also varied depending on the optics setting.
Typical values for each setting are shown in Table~\ref{table:optics}.
When the beta functions are squeezed, the collimator apertures are reduced accordingly.
Since the horizontal beam size is generally larger than the vertical size, the horizontal collimators are opened wider.

\begin{table}[hbt]
\centering
\begin{tabular}{cc|cccc}
\hline
                                                                            &                & D02V1 & D02H1 & D06V1 & D06H4 \\ \hline
\multicolumn{1}{c|}{\multirow{2}{*}{Detuned optics}}                        & (mm)           & 13    & 20    & 9     & 30    \\ \cline{2-6} 
\multicolumn{1}{c|}{}                                                       & in beam sigmas & 450   & 80    & 130   & 110   \\ \hline
\multicolumn{1}{c|}{\multirow{2}{*}{$\beta_y^{\ast}=1~\mathrm{mm}$ optics}} & (mm)           & 2     & 14    & 4     & 20    \\ \cline{2-6} 
\multicolumn{1}{c|}{}                                                       & in beam sigmas & 70    & 50    & 60    & 70    \\ \hline
\end{tabular}

\caption{Typical collimator apertures for the two main optics settings of the SuperKEKB main ring. The collimator names correspond to those shown in Fig.~\ref{fig:collimator_location}. The values represent the distance between opposing collimator heads. Both the physical distance (in mm) and the corresponding value normalized to the local beam size (in units of $\sigma$) are shown. In practice, these apertures are fine-tuned around the listed values during operation. The normalization to beam size was calculated using typical beam sizes in 2024 of $\sigma_x = 120~\mu$m and $\sigma_y = 70~\mu$m at the X-ray beam size monitor~\cite{Mulyani:2019gsy} location ($\beta_x = 4.57$ m, $\beta_y = 69.1$ m)~\cite{Ohnishi:eeFACT25}.}
\label{table:optics}
\end{table}

The BORs are connected to the beam abort kicker trigger (see Section~\ref{lossmonitor_and_beamabortsystem}).
Upon receiving the trigger, the BOR halts its ring buffer and stores the recorded data.
This allows capture of the beam position and charge of all bunches in the main ring for 100 turns prior to the beam abort~\cite{Nomaru:2024qls}.

\section{BOR data analysis for SBL events}
\label{02_BOR_data_analysis_for_SBL_events}

In this section, we present the results of our analysis of bunch charge and bunch position oscillations during SBL events, using data recorded by the two BORs.

\subsection{Example of SBL observations}
\begin{figure}[hbt]
\centering
\includegraphics[width=0.7\columnwidth]{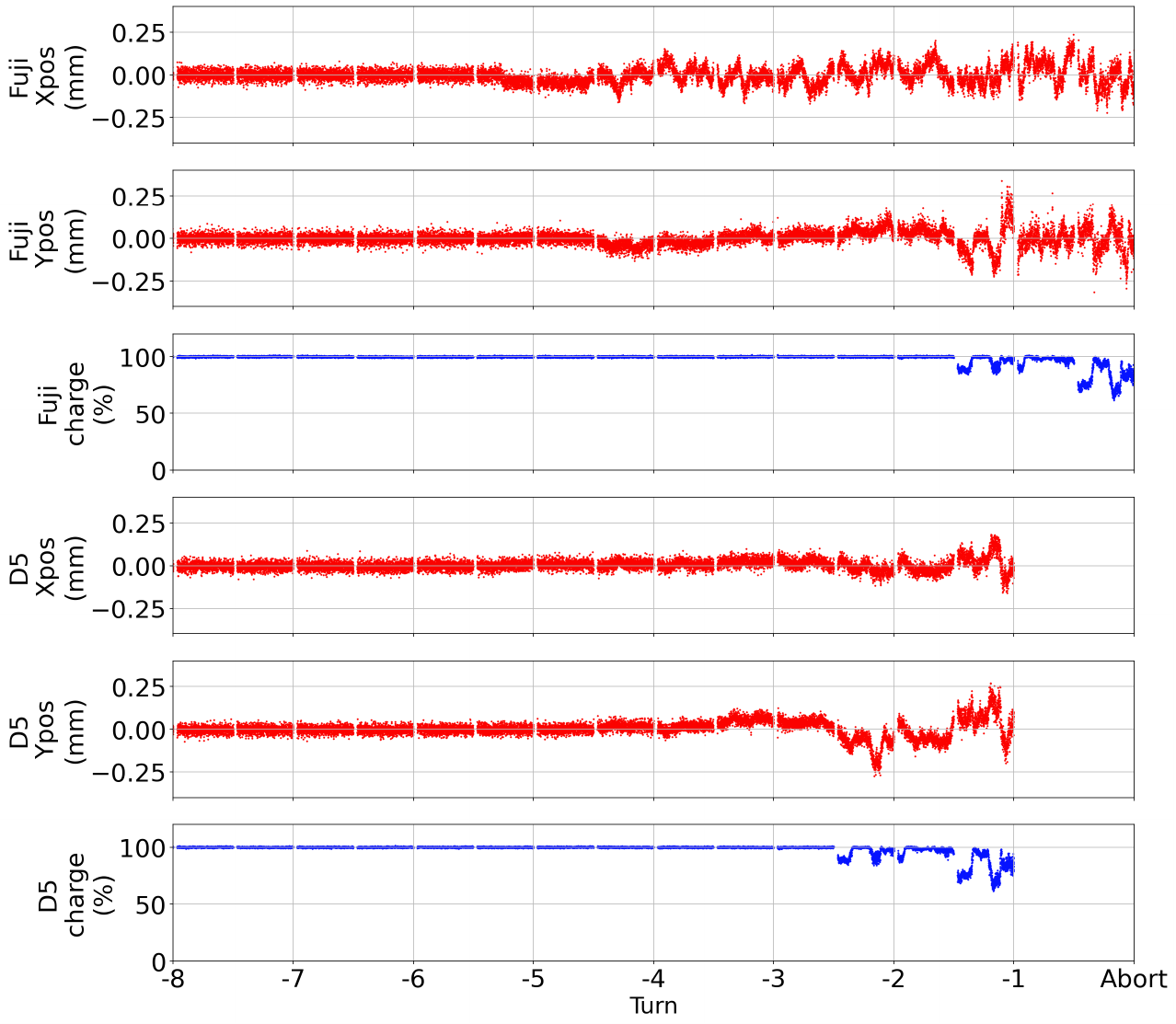}
\caption{An SBL event recorded by the Fuji-RFSoC (top three panels) and the D5-RFSoC (bottom three panels). The plots show bunch position and charge over the eight turns preceding the beam abort. From top to bottom: horizontal (X) position, vertical (Y) position, and bunch charge recorded by the Fuji-RFSoC, followed by the same measurements from the D5-RFSoC.}
\label{fig:sbl_10-28-025530}
\end{figure}

Figure~\ref{fig:sbl_10-28-025530} shows a representative example of an SBL event.
This event occurred at 2:55:30 on October 28, 2024, with 2346 bunches circulating in the ring and an average bunch current of 0.64~mA.
In this plot, each vertical division corresponds to one beam revolution (i.e., approximately 10~µs), and each dot represents a single bunch.
The BOR records each bunch once per turn, so the same bunch appears repeatedly in the plot.
The horizontal axis indicates the number of turns before the beam abort, with the abort occurring at the right edge of the plot.
Note that in the last turn before the abort, bunches pass through the location of the Fuji-RFSoC and then enter the beam dump (see Fig.~\ref{fig:collimator_location}).
Therefore, the final turn data is missing from the D5-RFSoC plot since it is located downstream of the beam dump.
In Fig.~\ref{fig:sbl_10-28-025530}, two bunch trains and two abort gaps can be identified within each revolution.
To isolate the oscillatory component due to beam instability leading to SBL, we subtract the bunch positions recorded in the first turn (100 turns before the abort) from the position of each bunch in all turns.
This removes offsets due to the nominal beam orbit.
For bunch charge, we normalize the values by defining the charge in the first recorded turn (100 turns before the abort) as 100\% for each bunch.
This normalization allows clearer visualization of charge loss because the bunch charges in the main ring are not perfectly uniform.
All subsequent plots in this paper follow the same conventions for representing bunch position and charge.
In this SBL event, oscillations in bunch positions began before any noticeable charge loss, both in the Fuji-RFSoC and D5-RFSoC data.
The beam abort was ultimately triggered with a maximum bunch charge loss of approximately 40\%.
As shown in Table~\ref{table:beta_nu}, the horizontal beta function at the D5-RFSoC installation location is relatively small, so the horizontal position oscillations of D5-RFSoC appears to be small in this plot.
Some oscillating bunches do not exhibit charge loss. Charge loss occurs when a portion of the beam hits either the collimator and the beam chamber, and therefore depends not only on the position oscillation but also on the beam size, oscillation amplitude, and phase. Consequently, it is trivial that some oscillating bunches do not experience charge loss.

\subsection{Bunch charge loss}
\label{bunch_charge_loss}
\begin{figure}[hbt]
 \centering
 \includegraphics[width=0.7\columnwidth]{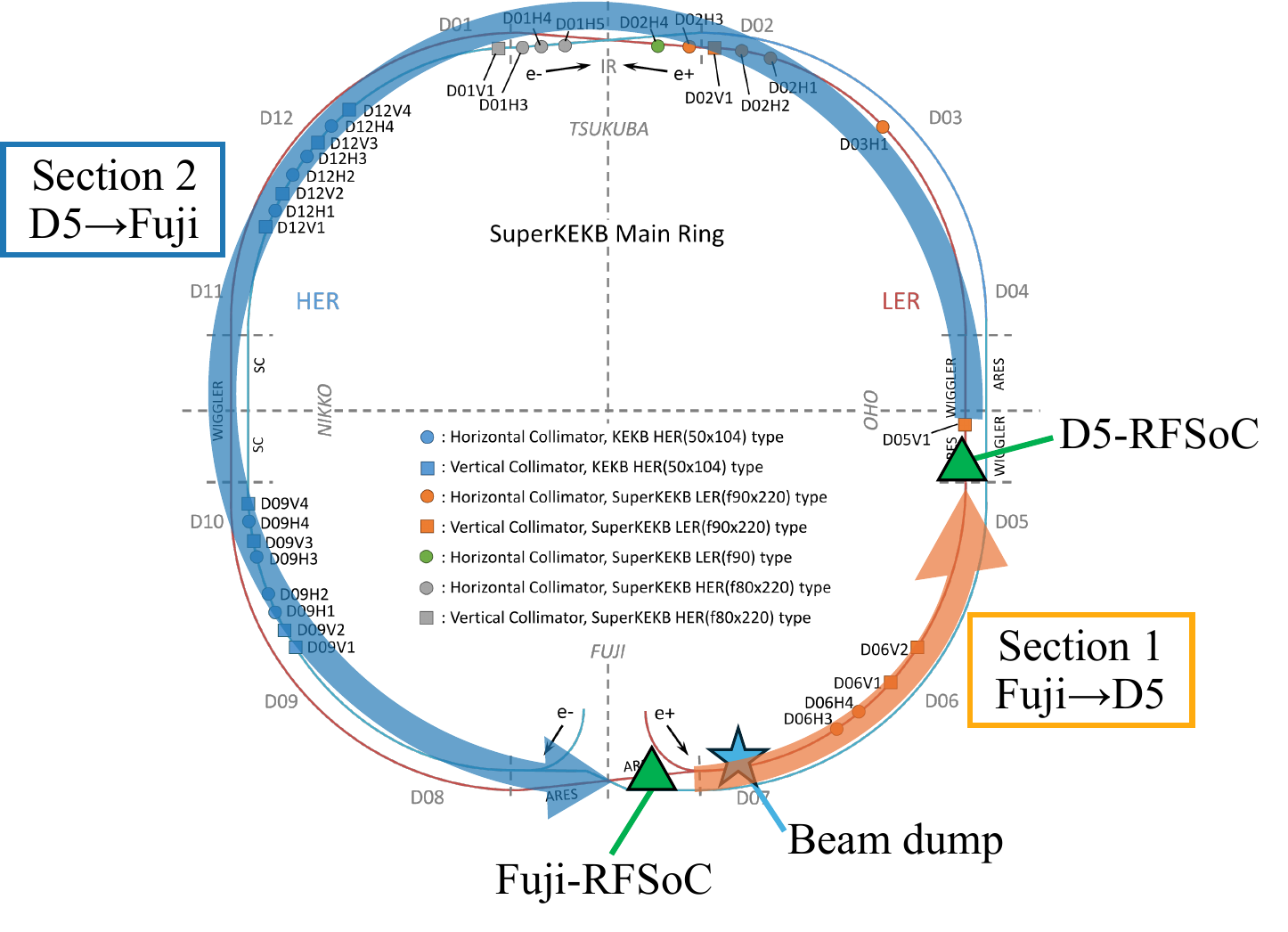}
 \caption{Division of the LER into two sections using the Fuji-RFSoC and the D5-RFSoC. Section~1 spans from the Fuji-RFSoC to the D5-RFSoC (orange arrow), and Section~2 spans from D5-RFSoC back to Fuji-RFSoC (blue arrow).}
 \label{fig:divide_sector}
\end{figure}

By combining data recorded by the Fuji-RFSoC and the D5-RFSoC, we investigate where in the main ring the charge loss occurs.  
As illustrated in Fig.~\ref{fig:divide_sector}, we conceptually divide the LER into two sections (“Section~1” and “Section~2”) using the two BORs.  
By comparing the bunch charge recorded by the Fuji-RFSoC with that recorded immediately afterward by the D5-RFSoC, we can determine the charge lost in Section~1.  
Similarly, by comparing the bunch charge recorded by the D5-RFSoC with that recorded by the Fuji-RFSoC one revolution later, we obtain the charge lost in Section~2.

\begin{figure}[hbt]
 \centering
 \includegraphics[width=0.8\columnwidth]{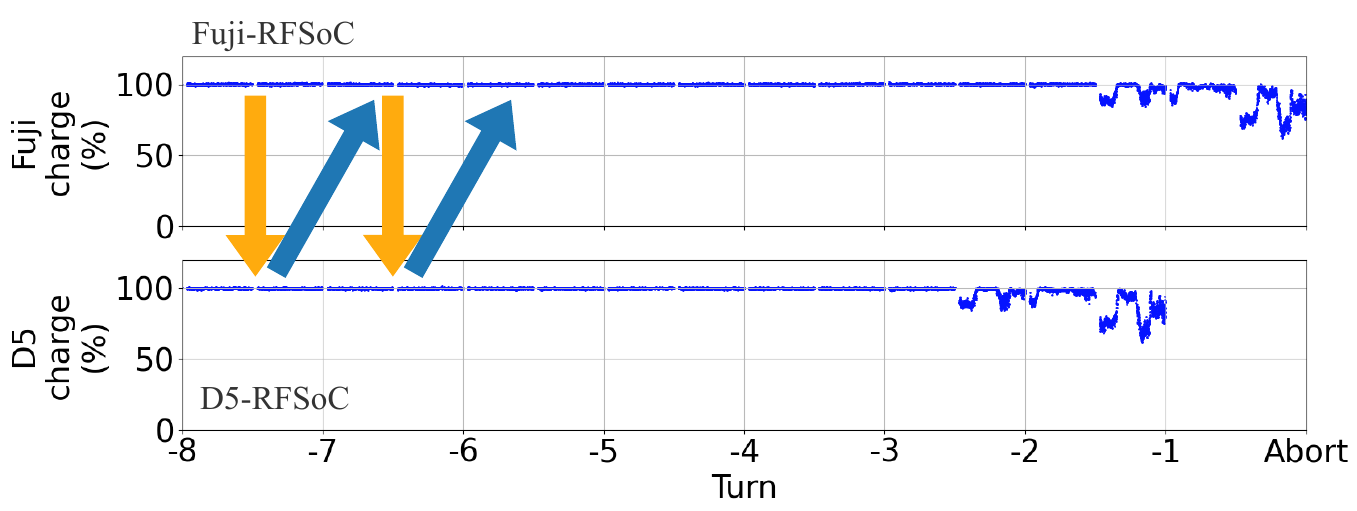}
 \caption{Bunch charge evolution during the SBL event at 02:55:30 on October 28, 2024. The upper panel shows data from the Fuji-RFSoC, and the lower panel shows data from the D5-RFSoC.}
 \label{fig:sbl_10-28-0235530_charge}
\end{figure}

As an example, Fig.~\ref{fig:sbl_10-28-0235530_charge} shows the bunch charge evolution for the same SBL event presented in Fig.~\ref{fig:sbl_10-28-025530}.  
In terms of time evolution, a bunch recorded by the Fuji-RFSoC moves through Section~1 and is then recorded by the D5-RFSoC—this corresponds to the transition from a point in the upper plot to the same horizontal position in the lower plot (orange arrows in Fig.~\ref{fig:sbl_10-28-0235530_charge}).  
Next, the bunch continues through Section~2 and returns to the Fuji-RFSoC one turn later, corresponding to the transition from a point in the lower plot to a point one turn later in the upper plot (blue arrows).  
These arrow colors match those in Fig.~\ref{fig:divide_sector}.
Since the beam passes through Section~2 just before the abort and is then dumped immediately after being recorded by the Fuji-RFSoC, the final turn is not recorded by the D5-RFSoC and appears blank.  
By comparing the upper and lower plots, we see that nearly all the charge loss occurred in Section~1, as indicated by the decrease in charge during the orange-arrow transitions.

\begin{figure}[hbt]
 \centering
 \includegraphics[width=0.8\columnwidth]{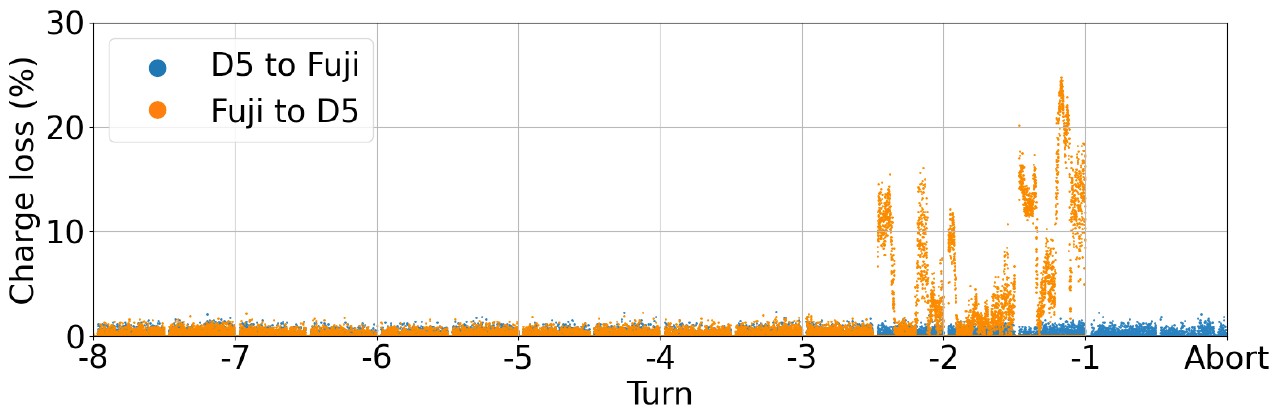}
 \caption{Charge loss per Section for the SBL event shown in Fig.~\ref{fig:sbl_10-28-0235530_charge}. Orange points (labeled “Fuji to D5”) represent charge loss in Section~1, while blue points (“D5 to Fuji”) represent loss in Section~2. Data for Section~1 in the final turn are not available.}
 \label{fig:sbl_10-28-025530_chargeloss}
\end{figure}

Figure~\ref{fig:sbl_10-28-025530_chargeloss} shows the result of subtracting the corresponding data points in the upper and lower plots of Fig.~\ref{fig:sbl_10-28-0235530_charge}.  
The orange points represent the difference between the Fuji-RFSoC and the D5-RFSoC in the same turn (i.e., the orange-arrow transitions), corresponding to charge loss in Section~1.  
The blue points represent the difference between the D5-RFSoC and the Fuji-RFSoC in the next turn (i.e., the blue-arrow transitions), corresponding to charge loss in Section~2.  
From this figure, we can conclude that almost all of the charge loss in this SBL event occurred in Section~1.  
In particular, this suggests that the beam hit collimators with narrow physical apertures in the D06 section, resulting in the charge loss.

\begin{figure}[hbt]
 \centering
 \includegraphics[width=0.45\columnwidth]{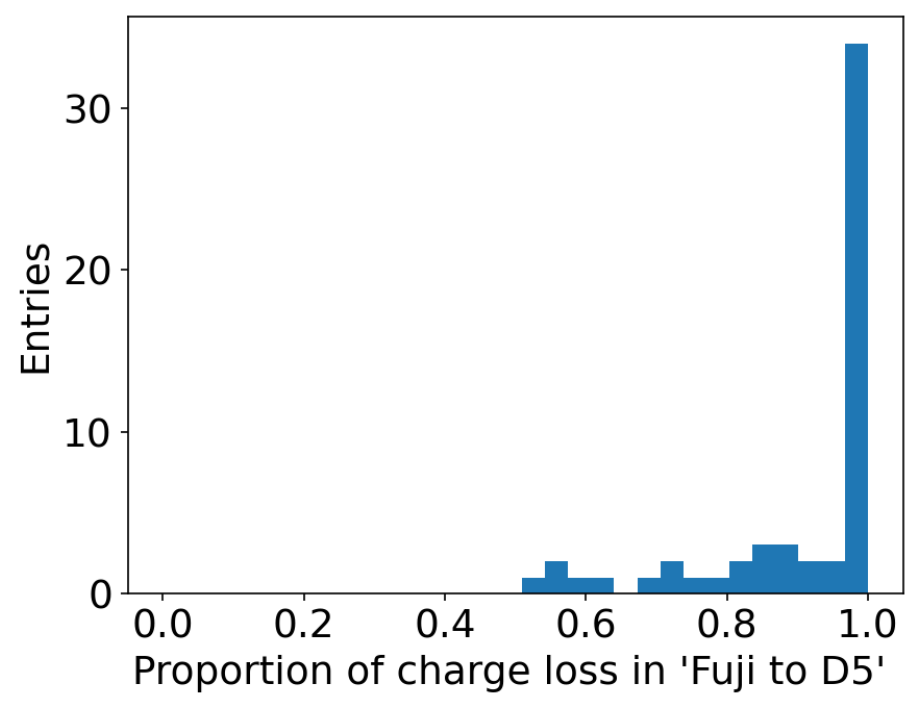}
 \caption{Distribution of the ratio \( P \) (defined in Eq.~\ref{eq:loss_percentage}) for 58 SBL events recorded simultaneously by the two BORs.}
 \label{fig:percentage_loss}
\end{figure}

To compare the distribution of charge loss between the two sections across multiple SBL events, we define the ratio \( P \) as follows:
\begin{equation}
    P = \frac{\text{Total charge loss in Section~1}}{\text{Total charge loss in Section~1} + \text{Total charge loss in Section~2}} .
    \label{eq:loss_percentage}
\end{equation}
The total charge loss in Section~1 is calculated by summing the orange points in Fig.~\ref{fig:sbl_10-28-025530_chargeloss} from 8 turns before the abort up to 1 turn before.  
For Section~2, the total loss is calculated by summing the blue points from 7 turns before the abort up to the final recorded turn.
Figure~\ref{fig:percentage_loss} shows a histogram of the ratio \( P \) calculated for 58 SBL events observed simultaneously by both BORs. 
In all cases, more than 50\% of the charge loss occurred in Section~1.  
This indicates that the majority of SBL-related charge loss can be attributed to beam interaction with collimators in the D06 section.
Among the many collimators installed in the ring, the analysis indicates that not those located in the D02 or D05 sections but specifically the collimator in the D06 section plays a critical role during SBL events.
This, in turn, implies that large position oscillations or possibly an increase in beam size tend to occur before the beam enters the D06 section.

\subsection{Duration of bunch position oscillation}
\label{oscillation_duration}
We next analyze the duration of bunch position oscillations.
In this analysis, we identify the onset time of oscillations and evaluate the duration until the beam abort.
To detect the onset time, we apply a moving average method to the position data.
We compute the average of the positions of 500 consecutive bunches, and then obtain a moving average by sliding the computed range one bunch at a time.
The onset of oscillation is defined as the moment when the absolute value of the moving average exceeds a predefined threshold.

\begin{figure}[hbt]
 \centering
 \includegraphics[width=0.45\columnwidth]{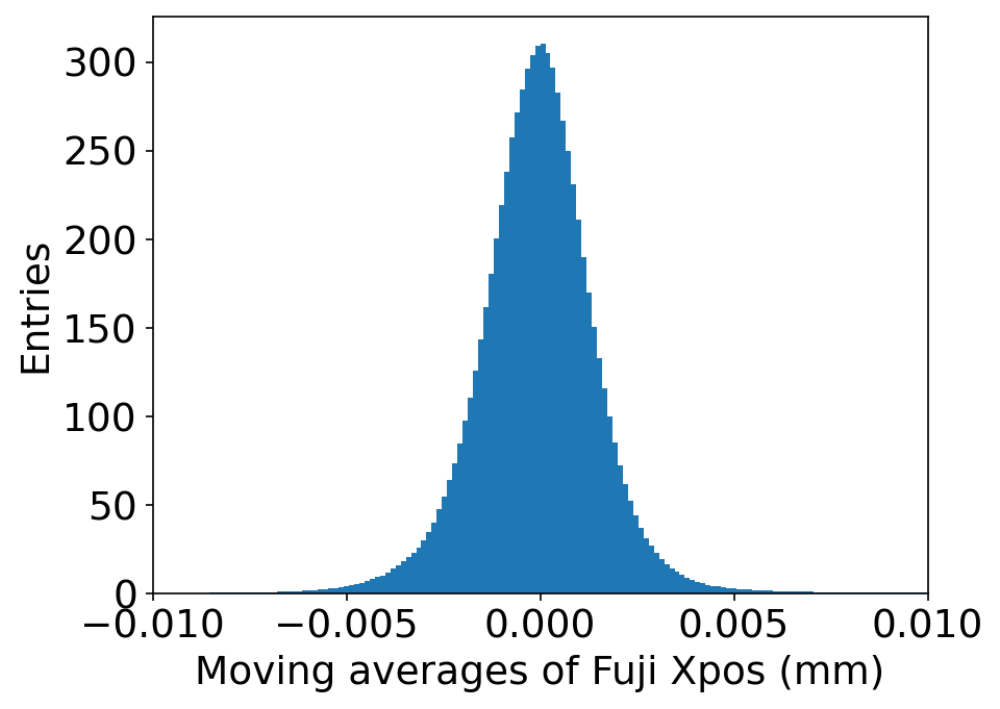}
 \caption{Histogram of the moving average values of the horizontal (X) position at Fuji-RFSoC under stable beam conditions.}
 \label{fig:threshold_set}
\end{figure}

The threshold is determined based on the fluctuation of the moving average under stable beam conditions.
To quantify this fluctuation, we use bunch position data collected by triggering the BOR periodically under stable beam conditions, rather than during beam aborts.
As an example, Fig.~\ref{fig:threshold_set} shows the distribution of the moving average computed from approximately 30~ms of horizontal position data recorded by the Fuji-RFSoC during a stable period.
The resulting distribution is centered at zero with a standard deviation of $\sigma = 0.002~\mathrm{mm}$.
This is consistent with the expected fluctuation of the moving average, estimated from the BOR's position resolution of 30~$\mu$m~\cite{Nomaru:2024qls} as \(30~\mu\mathrm{m}/\sqrt{500} \sim 1.3~\mu\mathrm{m}\).
Based on this result, we set the threshold to \(5\sigma = 0.010~\mathrm{mm}\).
Using the same procedure, thresholds are determined for both horizontal and vertical positions at the Fuji-RFSoC and the D5-RFSoC, as summarized in Table~\ref{table:threshold_setting}.

\begin{table}[h]
\centering
\begin{tabular}{c|cccc}
\hline
        & \begin{tabular}[c]{@{}c@{}}Fuji-RFSoC\\ X position\end{tabular} & \begin{tabular}[c]{@{}c@{}}Fuji-RFSoC\\ Y position\end{tabular} & \begin{tabular}[c]{@{}c@{}}D5-RFSoC\\ X position\end{tabular} & \begin{tabular}[c]{@{}c@{}}D5-RFSoC\\ Y position\end{tabular} \\ \hline
Threshold (mm) & 0.010 & 0.012 & 0.012 & 0.014 \\ \hline
\end{tabular}
\caption{Threshold values used to detect the onset of oscillations in the horizontal (X) and vertical (Y) position for each BOR.}
\label{table:threshold_setting}
\end{table}

\begin{figure}[hbt]
 \centering
 \includegraphics[width=0.7\columnwidth]{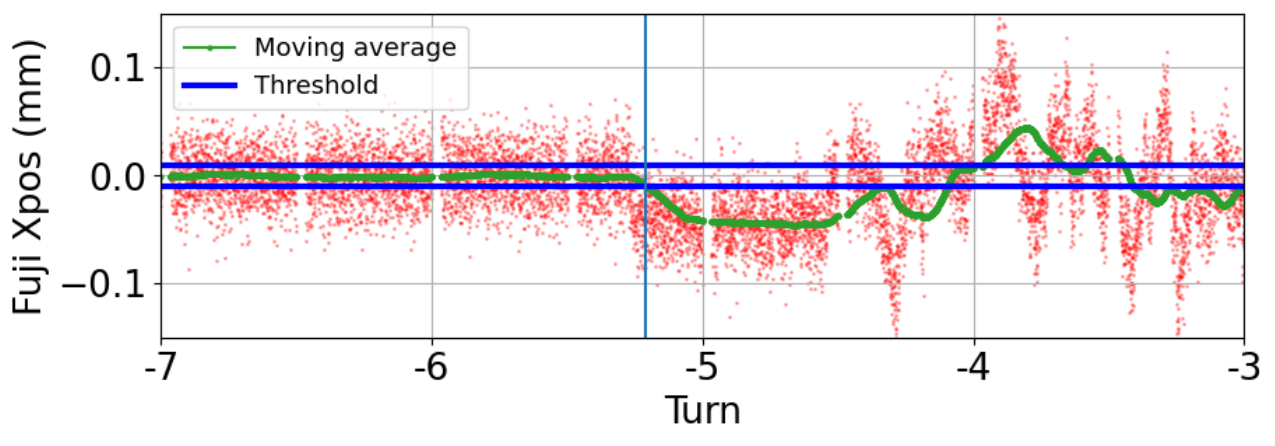}
 \caption{Enlarged view of the top panel in Fig.~\ref{fig:sbl_10-28-025530}, showing the horizontal (X) position data from the Fuji-RFSoC. Red points indicate bunch positions, the green line is the moving average, and the blue horizontal lines represent the thresholds. The value of the moving average is plotted at the location corresponding to the last data point used in its calculation. The vertical blue line marks the detected onset of oscillation.}
 \label{fig:sbl_10-21-085935_ex}
\end{figure}

Figure~\ref{fig:sbl_10-21-085935_ex} illustrates an example of how oscillation onset is detected.
The onset is defined as the point where the moving average exceeds either the positive or negative threshold (blue horizontal lines).
In this example, the horizontal position oscillation is found to begin 5.2 turns before the beam abort.

\begin{figure}[hbt]
 \centering
 \begin{subfigure}[b]{0.8\textwidth}
 \centering
 \includegraphics[width=1\columnwidth]{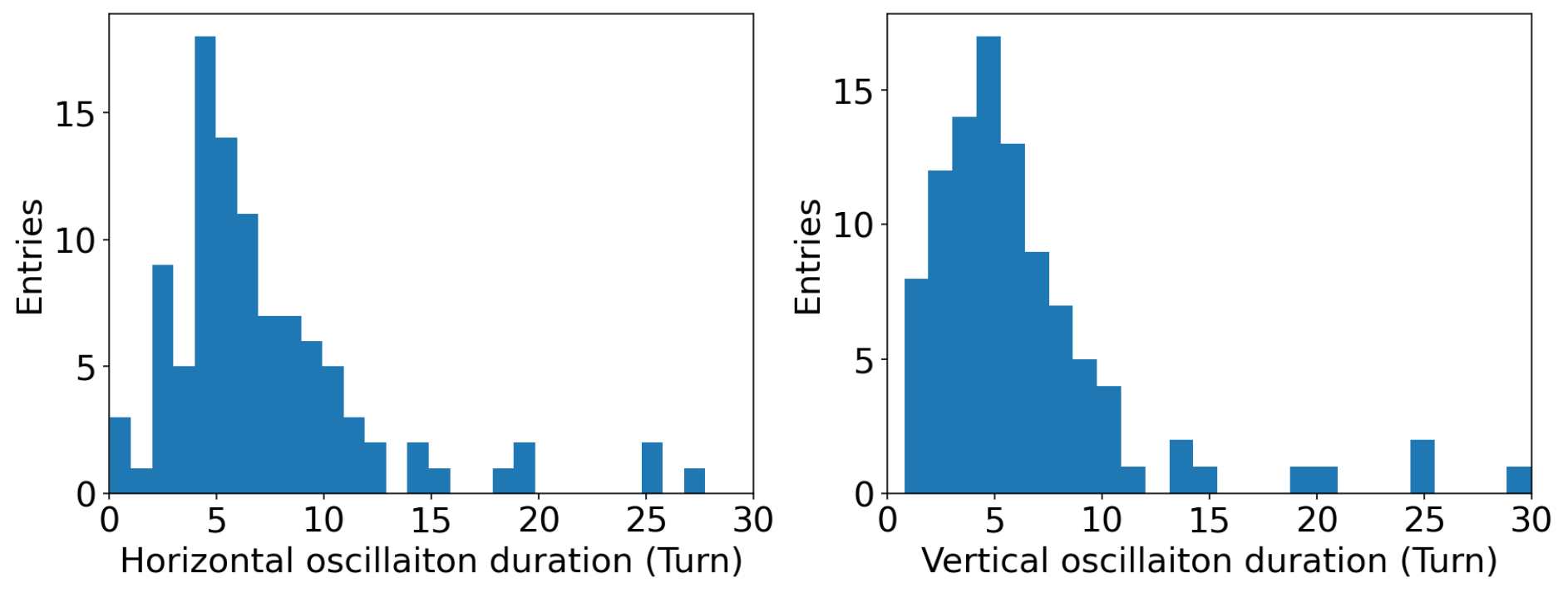}
 \subcaption{117 SBL events recorded by the Fuji-RFSoC}
 \label{fig:oscillation_duration_fuji}
\end{subfigure}
\hfill
 \begin{subfigure}[b]{0.8\textwidth}
 \centering
 \includegraphics[width=1\columnwidth]{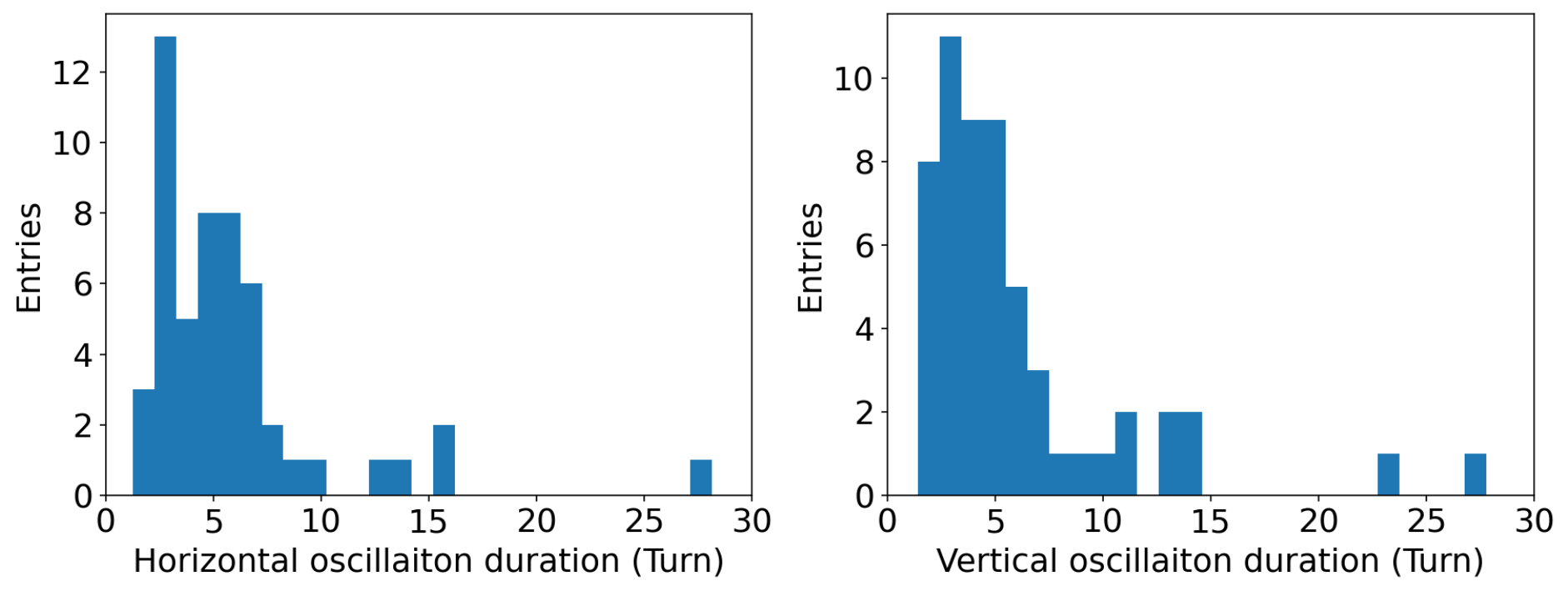}
 \subcaption{58 SBL events recorded by the D5-RFSoC}
 \label{fig:oscillation_duration_d5}
\end{subfigure}
\caption{Histograms showing the duration of oscillations prior to beam abort. Left: horizontal oscillations; Right: vertical oscillations. Only events with 30 turns or fewer are shown.}
\label{fig:oscillation_duration}
\end{figure}

We apply this method to all SBL events recorded by the Fuji-RFSoC and D5-RFSoC to determine the duration of oscillations prior to the beam abort.
The resulting distributions are shown in Fig.~\ref{fig:oscillation_duration}.
For the Fuji-RFSoC, 69\% of SBL events for horizontal oscillation and 73\% of SBL events for vertical oscillation have a duration of 10 turns or less.
For the D5-RFSoC, 81\% of SBL events for horizontal oscillation and 83\% of SBL events for vertical oscillation have a duration of 10 turns or less.
These results indicate that in the majority of SBL events, bunch position oscillations leading to beam aborts typically last only tens of microseconds, up to around 100~µs.

The duration of these oscillations is often longer than the interval between the onset of beam loss and the beam abort. 
If the onset of bunch position oscillations can be detected in real time and used as a trigger for the abort system, it may be possible to abort the beam before a large loss occurs, thereby protecting the accelerator from radiation damage. 
We are currently implementing and testing additional BOR functions for oscillation detection based on this moving average method and for trigger generation~\cite{nomaru:ipac2025-thpm089}.
In the future, the BOR may also serve as part of the accelerator protection system.

\subsection{Amplitude of Bunch Position Oscillation}
\label{Oscillation_amplitude}
\begin{figure}[hbt]
 \centering
 \includegraphics[width=0.8\columnwidth]{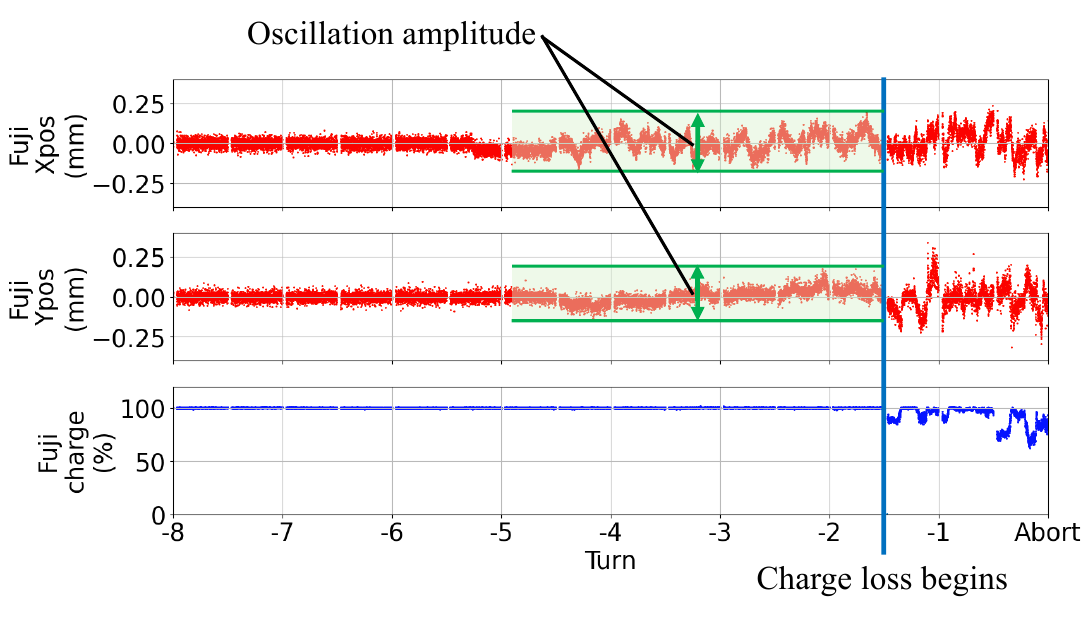}
 \caption{Schematic of the amplitude calculation, based on the top three panels of Fig.~\ref{fig:sbl_10-28-025530}. The amplitude is defined as the difference between the maximum and minimum bunch positions before this point. The blue vertical line marks the onset of bunch charge loss.}
 \label{fig:shinpuku_gaiyou}
\end{figure}

\begin{figure}[hbt]
 \centering
 \begin{subfigure}[b]{0.8\textwidth}
 \includegraphics[width=1\columnwidth]{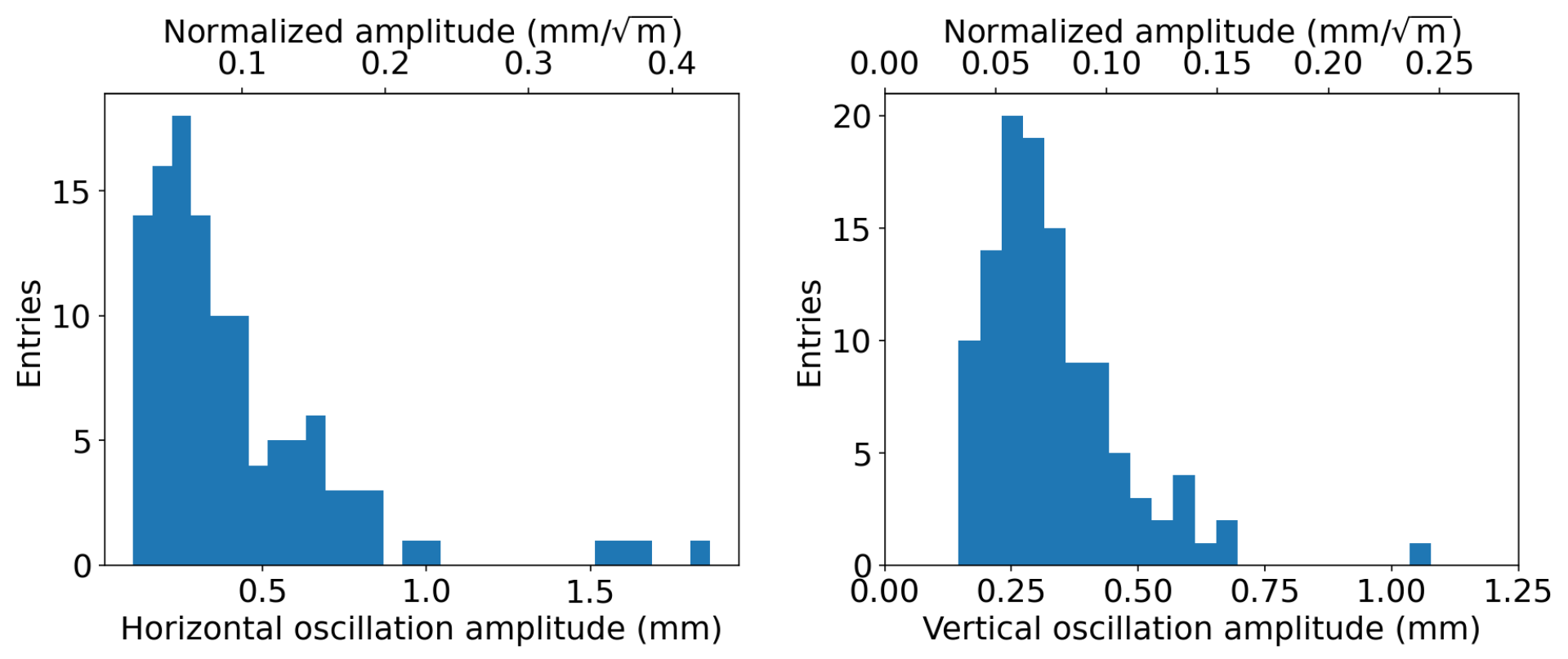}
 \subcaption{117 SBL events recorded by the Fuji-RFSoC}
 \label{fig:oscillation_amp_fuji}
\end{subfigure}
\hfill
\begin{subfigure}[b]{0.8\textwidth}
 \centering
 \includegraphics[width=1\columnwidth]{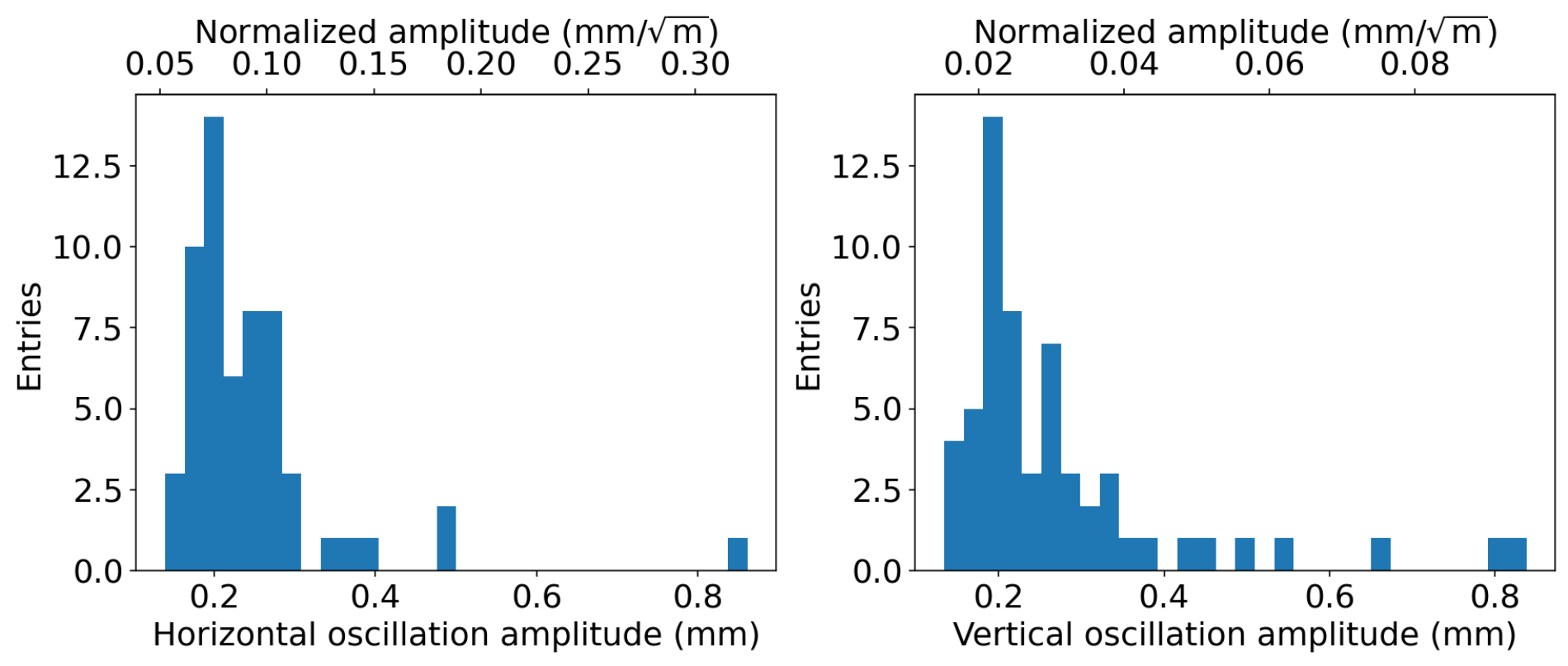}
 \subcaption{58 SBL events recorded by the D5-RFSoC}
 \label{fig:oscillation_amp_d5}
\end{subfigure}
\caption{Distributions of bunch oscillation amplitude. Left: horizontal oscillation. Right: vertical oscillation. The top axis in each plot shows the normalized amplitude, obtained by dividing the amplitude (in mm) by the square root of the beta function ($\sqrt{\mathrm{m}}$) at the BOR location.}
\label{fig:oscillation_amp}
\end{figure}

We next investigate the amplitude of bunch position oscillations during SBL events.  
Figure~\ref{fig:shinpuku_gaiyou} illustrates the procedure used to determine the oscillation amplitude.  
Bunches that have already lost charge are excluded from the calculation, since partial loss due to interactions with collimators can shift the center of charge and distort the position measurement.
Here, we define the amplitude as the difference between the maximum and minimum values of the bunch position recorded before the onset of charge loss.
The onset of charge loss is defined as the moment when the charge of any bunch decreases by more than 5\%.  
Using this definition, we calculate the oscillation amplitudes for all SBL events recorded by the Fuji-RFSoC and the D5-RFSoC.
The resulting distributions are shown in Fig.~\ref{fig:oscillation_amp}.
In addition to the raw amplitude (in mm), we present the “normalized amplitude,” defined as the amplitude divided by the square root of the beta function ($\sqrt{\beta_{x,y}(\mathrm{m})}$) at the BOR location.  
This normalization provides a more intrinsic measure of oscillation strength that is independent of the observation point.  
The normalized amplitudes are displayed on the top axes of the histograms.  

Focusing on the normalized amplitudes in the vertical oscillation, the results indicate that the Fuji-RFSoC tends to exhibit larger normalized amplitudes than the D5-RFSoC.
This difference likely arises from our use of only pre-charge-loss data in the amplitude calculation.  
When strong bunch oscillations are observed at the Fuji-RFSoC, similar amplitudes would be expected at the D5-RFSoC under normal conditions.  
However, if beam loss occurs at the D06 collimator section immediately downstream of the Fuji-RFSoC, those bunches are no longer included in the amplitude calculation at the D5-RFSoC.  
As a result, the calculated amplitudes at the D5-RFSoC tend to be smaller than those at the Fuji-RFSoC.
This observation suggests that strong bunch oscillations, or possibly an increase in beam size, often develop before the beam enters the D06 section and hit the collimators there—consistent with the discussion in Section~\ref{bunch_charge_loss}.

\section{Analysis Using Multiple Monitoring Systems}
\label{03_Analysis_Using_Multiple_Monitoring_Systems}

In this section, we incorporate information from vacuum gauges and loss monitors into the BOR-based analysis to gain deeper insight into the causes of SBL events.
In this section, each CCG is referred to by its location name only. 
For example, “D01\_L01 CCG” is simply denoted as “D01\_L01.”

\subsection{Relationship between pressure burst locations and bunch oscillations}
\begin{figure}[hbt]
 \centering
 \includegraphics[width=0.6\columnwidth]{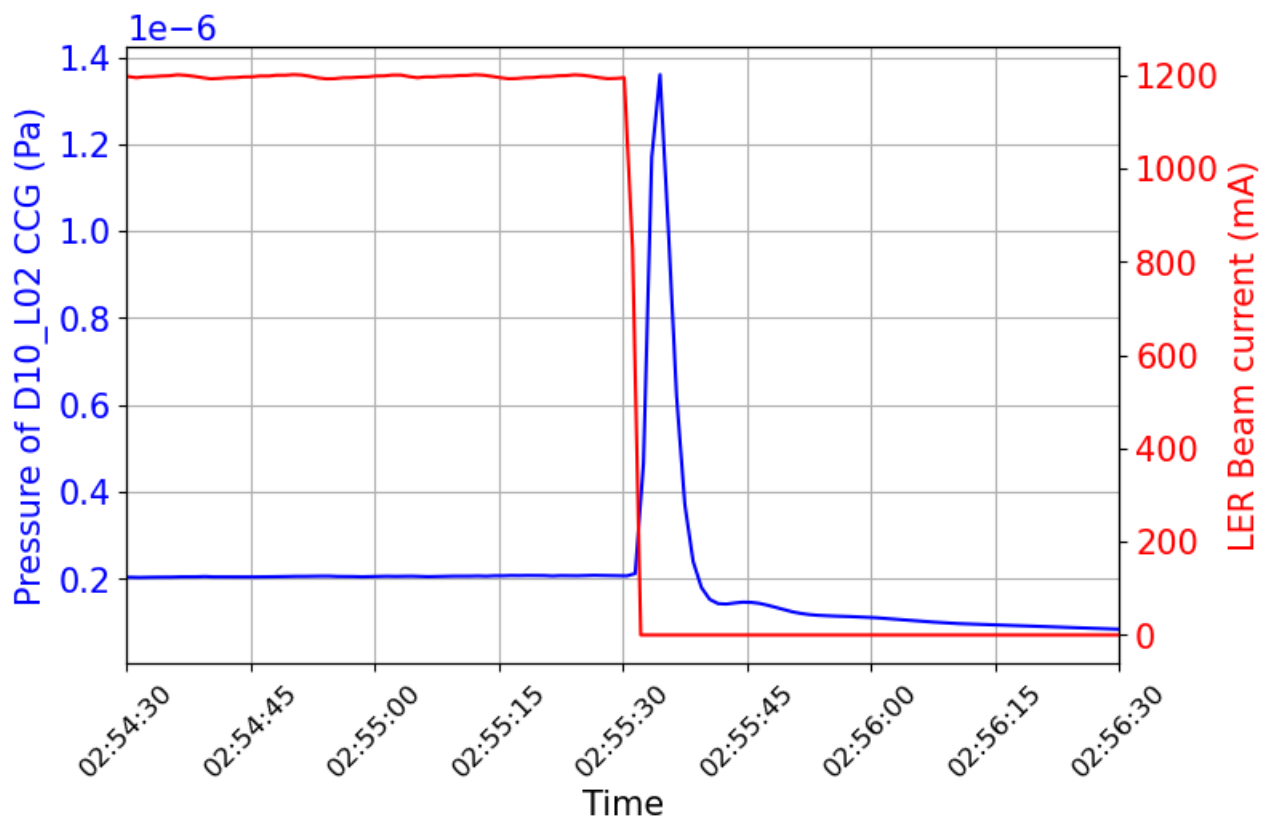}
 \caption{LER beam current (red) and vacuum pressure (blue) measured by the CCG at D10\_L02 during the SBL event at 2:55:30 on October 28, 2024.}
 \label{fig:pressureburst_example}
\end{figure}

We begin by describing the phenomenon of pressure bursts~\cite{Terui:2018pae}, which are suspected to be closely related to SBL events.
During SBL events, a sudden and abnormal increase in vacuum pressure inside the beam chamber is frequently observed at certain points in the main ring.
As an example, Fig.~\ref{fig:pressureburst_example} shows the time evolution of the vacuum pressure.
It was measured by the CCG installed at D10\_L02 during the SBL event shown in Fig.~\ref{fig:sbl_10-28-025530}.
It can be seen that the vacuum pressure rises sharply after the beam current drops to zero due to the beam abort.
Because the response time of the CCG is relatively slow, it is believed that the pressure spike occurs either simultaneously with or just before the beam abort.
We refer to such phenomena as “pressure bursts.”
Note that, in general, pressure bursts are not observed during controlled, safe beam aborts that do not involve beam loss.

To ensure consistent beam conditions for comparison, we focus on 58 SBL events recorded by both BORs during the period from October 24 to November 23, when SuperKEKB was operated under the \(\beta_y^{\ast} = 1~\mathrm{mm}\) optics.
Of these, 55 events were accompanied by pressure bursts.
Table~\ref{table:pressure_burst_table} lists the locations and number of occurrences of pressure bursts.
As shown in the table, the majority of pressure bursts during this period occurred in the D10 section.
Within D10, several distinct CCGs recorded pressure burst events, with the combination of D10\_L02 and D10\_L03 accounting for 60\% of all cases.
These two CCGs are located adjacent to each other and consistently detected pressure bursts simultaneously, so they are treated as a single location in the table.

\begin{table}[hbt]
\centering
\begin{tabular}{c|c}
\hline
CCG Location where\\pressure burst was observed & Number of Events \\ \hline
D10\_L02/03   & 33 \\
D10\_L05      & 4  \\
D10\_L06      & 2  \\
D10\_L07      & 2  \\
D10\_L08      & 2  \\
D02\_L18      & 2  \\
D06\_L12      & 2  \\ 
\hline
\end{tabular}
\caption{Locations where pressure bursts were observed during SBL events, listed in descending order of frequency. Only locations with two or more occurrences are shown. D10\_L02/03 represents the combined detection from adjacent CCGs at D10\_L02 and D10\_L03, which always respond simultaneously.}
\label{table:pressure_burst_table}
\end{table}

\begin{figure}[hbt]
 \centering
 \begin{subfigure}[b]{0.8\textwidth}
 \includegraphics[width=1\columnwidth]{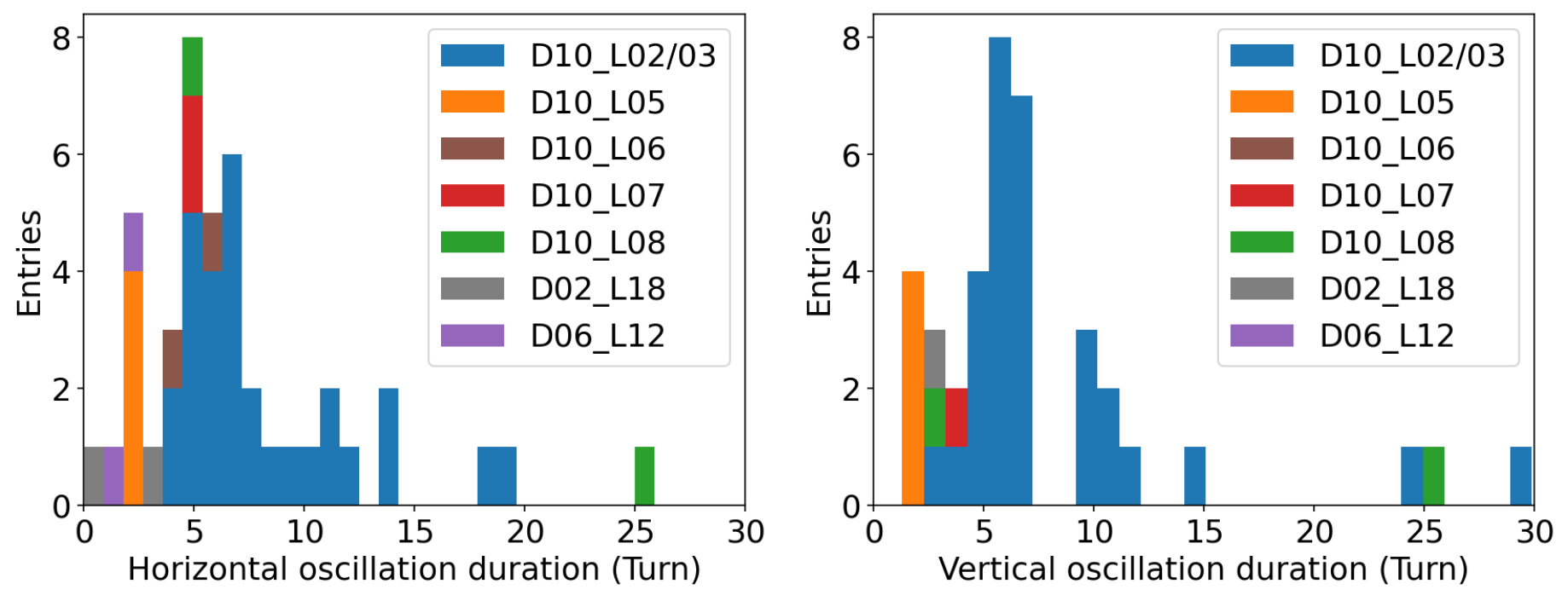}
 \subcaption{Fuji-RFSoC}
 \label{fig:duration_pressure_fuji}
\end{subfigure}
\hfill
\begin{subfigure}[b]{0.8\textwidth}
 \centering
 \includegraphics[width=1\columnwidth]{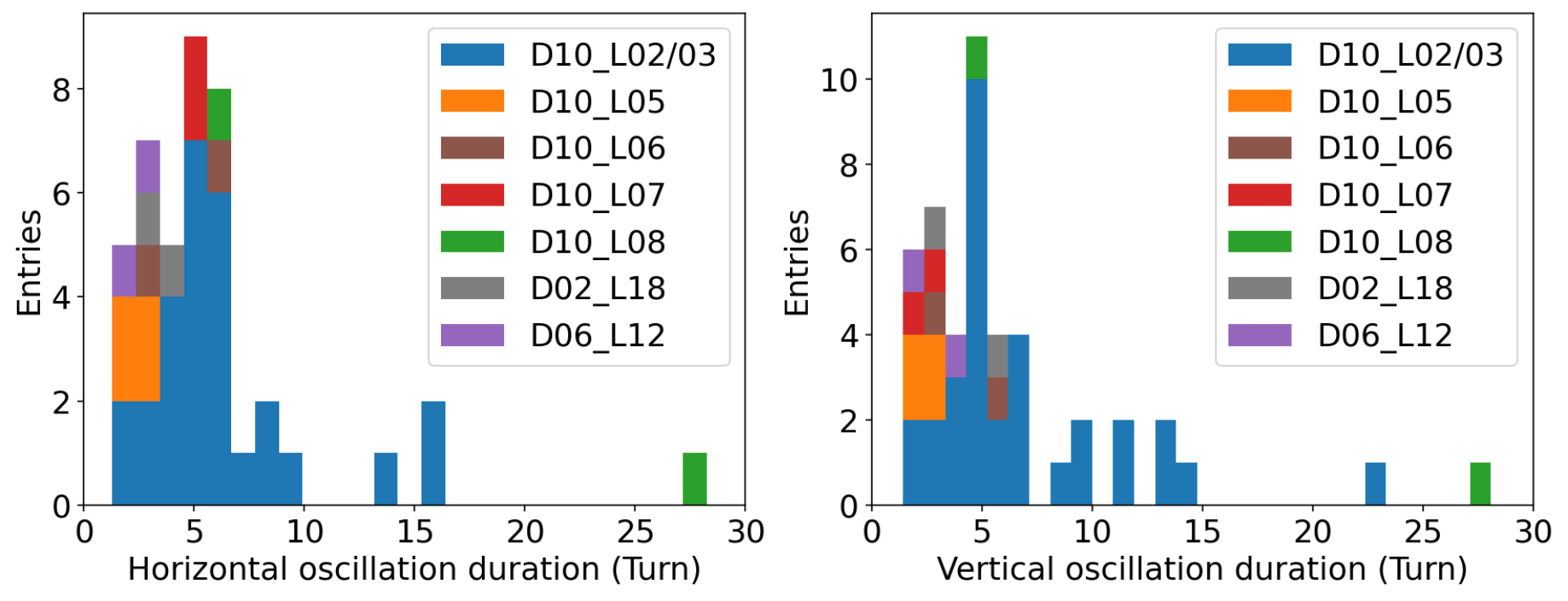}
 \subcaption{D5-RFSoC}
 \label{fig:duration_pressure_d5}
 \end{subfigure}
 \caption{Histograms of oscillation duration for SBL events accompanied by pressure bursts. Left: horizontal oscillations; Right: vertical oscillations. Histogram colors correspond to the pressure burst locations. Only events with oscillation durations of 30 turns or fewer are shown. The data in each color are stacked in the histogram.}
 \label{fig:duration_pressure}
\end{figure}

The duration of bunch oscillations prior to the beam abort is determined using the method described in Section~\ref{oscillation_duration}.
The results are categorized by pressure burst location, and the corresponding histograms are shown in Fig.~\ref{fig:duration_pressure}.
Focusing on the two most frequently observed pressure burst locations—D10\_L02/03 and D10\_L05—we find the following:
\begin{itemize}
    \item In all four histograms, SBL events associated with pressure bursts at D10\_L05 (orange) are concentrated in the short-duration region.
    \item In contrast, events associated with D10\_L02/03 (blue) tend to exhibit longer oscillation durations than D10\_L05.
\end{itemize}
In the following, we examine and discuss SBL events accompanied by pressure bursts at these two locations separately.

\subsection{Potential scenario of SBL evolution arising at D10\_L02/L03}
\label{section:D10L02/03}
\begin{figure}[hbt]
 \centering
 \includegraphics[width=0.6\columnwidth]{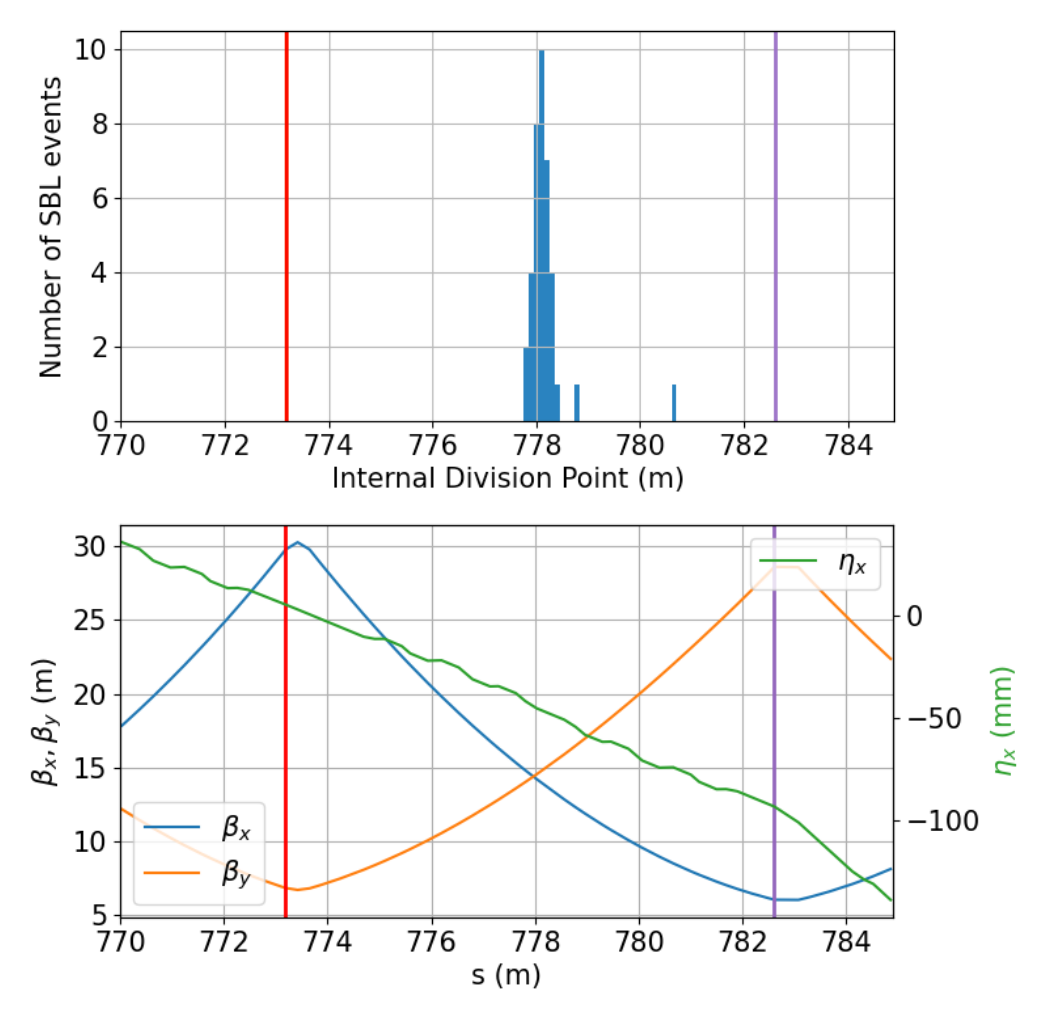}
 \caption{(Top) Estimated distribution of pressure burst origins based on pressure values observed at D10\_L02 and D10\_L03. The horizontal axis indicates the distance from the collision point. Red and purple vertical lines show the positions of D10\_L02 and D10\_L03, respectively. (Bottom) Beta and dispersion functions near these CCGs. The left vertical axis corresponds to the beta function, and the right to the horizontal dispersion function. The horizontal axes are aligned with the top figure.}
 \label{fig:lattice_D10_L02}
\end{figure}

D10\_L02 and D10\_L03 are the locations where pressure bursts accompanying SBL events were most frequently observed.  
Notably, pressure bursts were always detected simultaneously by the adjacent CCGs at D10\_L02 and D10\_L03.  
This suggests that the pressure burst originated between these two gauges and propagated outward, resulting in simultaneous detection.  
In previous studies during SuperKEKB operation in 2016, the location of pressure bursts was inferred from the pressure values recorded at surrounding CCGs~\cite{Terui:2018pae}.
Figure~\ref{fig:lattice_D10_L02}~(top) shows the estimated distribution of pressure burst origins, obtained by calculating the internal division point between the pressure values at D10\_L02 and D10\_L03 for 33 SBL events.  
Figure~\ref{fig:lattice_D10_L02} (bottom) shows the beta and dispersion functions near these CCGs.  
From the top panel, we infer that the pressure burst most likely originated near the midpoint between D10\_L02 and D10\_L03, suggesting that the beam may have interacted with some material at this location and was subjected to an external force.
A bellows chamber is installed in this region.

\begin{figure}[hbt]
 \centering
 \includegraphics[width=0.7\columnwidth]{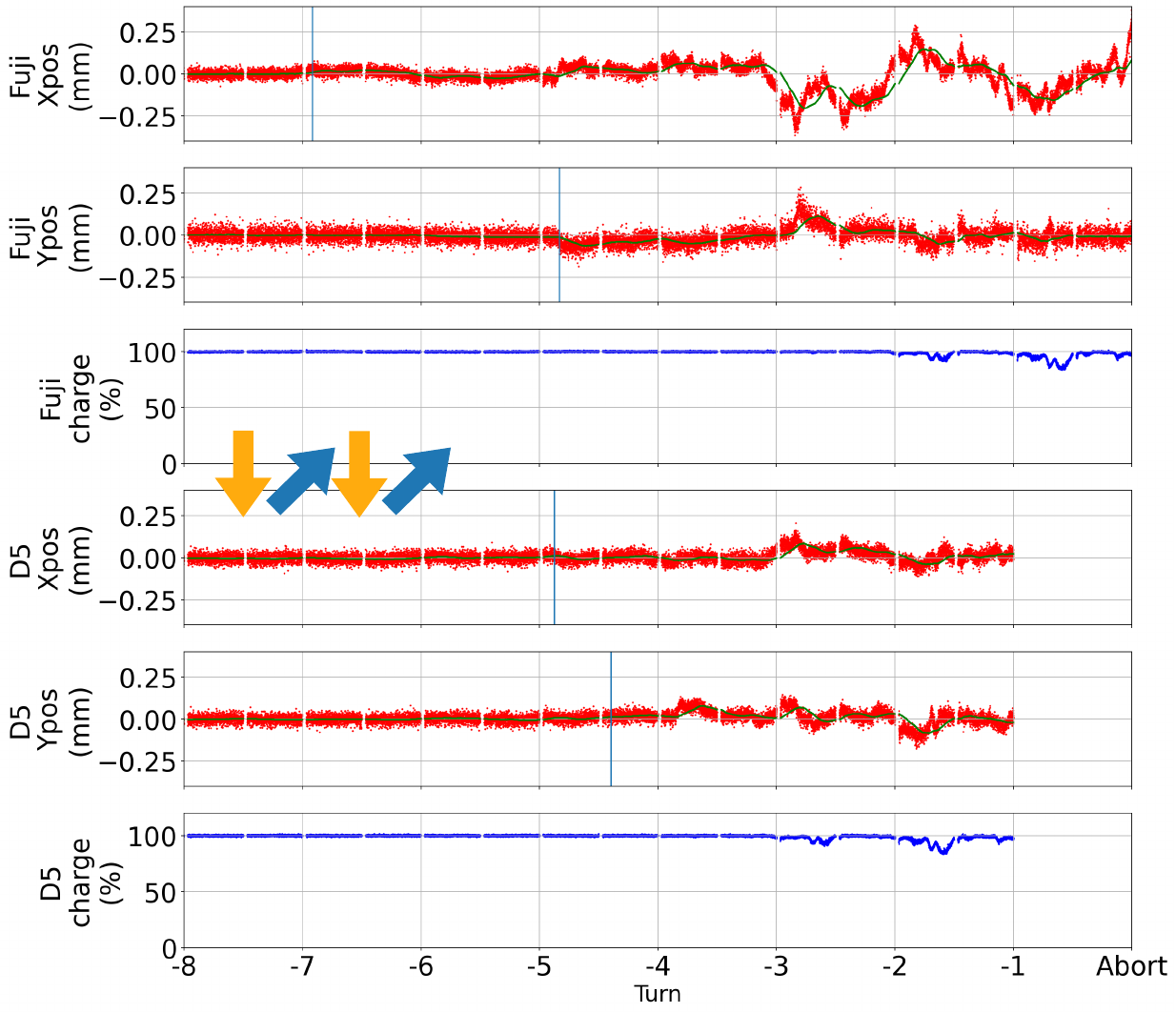}
 \caption{An SBL event accompanied by a pressure burst at D10\_L02/03. From top to bottom: Horizontal (X) position, Vertical (Y) position, and bunch charge recorded by the Fuji-RFSoC, followed by the same data from the D5-RFSoC. Vertical blue lines indicate the detected oscillation onset times. Green lines show the moving average traces.}
 \label{fig:sbl_1103_035826}
\end{figure}

Figure~\ref{fig:sbl_1103_035826} shows a typical SBL event associated with a pressure burst at D10\_L02/03, observed by the BORs.
This event occurred at 3:58:26 on November 3, 2024, with 2346 bunches circulating in the ring and an average bunch current of 0.45~mA.
The plot format is the same as in Fig.~\ref{fig:sbl_10-28-0235530_charge}, with data flowing downward from the Fuji-RFSoC (top three panels) to the D5-RFSoC (bottom three panels), corresponding to motion through Section~1 as shown in Fig.~\ref{fig:divide_sector} (orange arrows).
In this event, oscillations are first observed in the horizontal position at the Fuji-RFSoC (6.9 turns before the abort), followed by the horizontal position at the D5-RFSoC (4.9 turns), vertical position at the Fuji-RFSoC (4.8 turns), and finally vertical position at the D5-RFSoC (4.4 turns before the abort).

\begin{figure}[hbt]
 \centering
 \includegraphics[width=0.4\columnwidth]{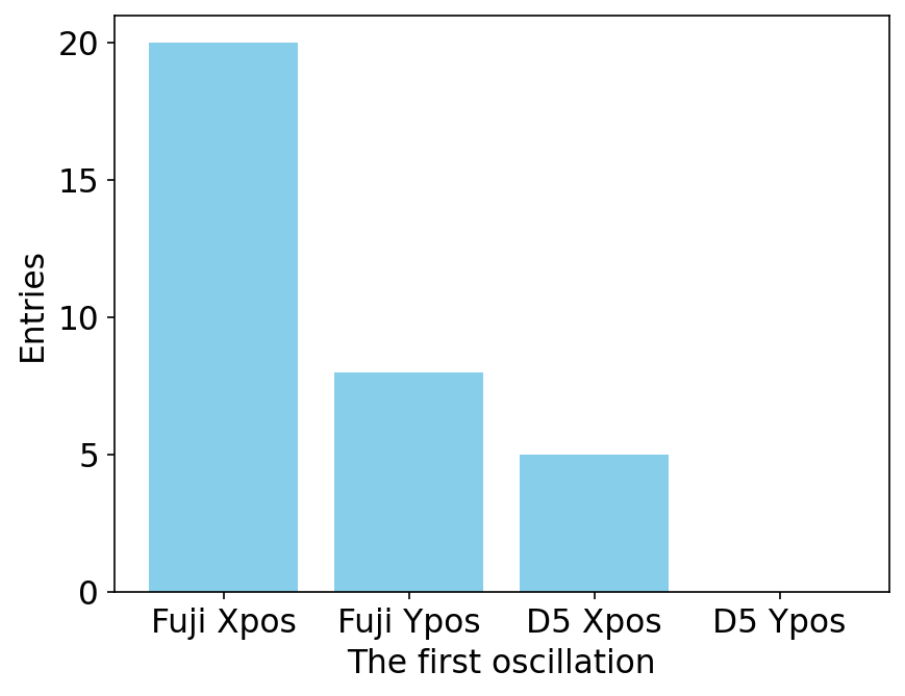}
 \caption{Number of SBL events with D10\_L02/03 pressure burst in which each signal—horizontal (X) or vertical (Y) position at Fuji-RFSoC or D5-RFSoC—exhibited the earliest onset of oscillation.}
 \label{fig:first_oscillation}
\end{figure}

Figure~\ref{fig:first_oscillation} summarizes, for the 33 SBL events accompanied by pressure bursts at D10\_L02/03, which signal (horizontal position or vertical position at the Fuji or D5-RFSoC) shows the earliest oscillation onset.  
The results show a clear tendency for oscillations to begin earlier at the Fuji-RFSoC than at the D5-RFSoC.

If the beam receives a kick in the D10 section, the resulting oscillation would propagate downstream, first passing the Fuji-RFSoC and then traveling through Section~1 to the D5-RFSoC.  
One might expect this to result in simultaneous observation of oscillation onset at both BORs in these plots.  
However, the observed amplitude at each location depends on the betatron phase relative to the kick location.  
This may delay the detection of the oscillation onset under conditions where position oscillation is not visible.
As inferred earlier, the beam likely received the kick force near the midpoint between D10\_L02 and D10\_L03.  
The bunch position observed at downstream “location~2” after receiving a kick at “location~1” can be expressed as:
\begin{equation}
    y_2 \approx \sqrt{\beta_{y1} \beta_{y2}} \sin \Psi_{12} \Delta y_1^{\prime},
    \label{eq:kick}
\end{equation}
where \(y_2\) is the bunch position at location 2, \(\beta_{y1}, \beta_{y2}\) are the beta functions at locations 1 and 2, \(\Psi_{12}\) is the betatron phase advance between location 1 and 2, and \(\Delta y_1^{\prime}\) is the kick angle.  
From this expression, the oscillation is most visible when \(\Psi_{12} \approx (n + 0.5)\pi\), and minimized when \(\Psi_{12} \approx n\pi\), where $n$ is an integer.
Table~\ref{table:D10_L02phase} summarizes the betatron phases of the Fuji-RFSoC and the D5-RFSoC relative to the midpoint between D10\_L02 and D10\_L03, based on Table~\ref{table:beta_nu}.
We find that the Fuji-RFSoC is approximately at a half-integer multiple of \(\pi\), while the D5-RFSoC is near an integer multiple for both horizontal and vertical directions.  
Thus, if a kick occurs at the midpoint, it is more likely to be strongly visible at the Fuji-RFSoC, consistent with the earlier onset detection there.

\begin{table}[h]
\begin{tabular}{c|ccc}
\hline
                   & Midpoint of D10\_L02 and D10\_L03 & Fuji-RFSoC & D5-RFSoC \\ \hline
Horizontal betatron phase [rad/\(\pi\)] & 0 & 21.40 & 40.89 \\
Vertical betatron phase [rad/\(\pi\)] & 0 & 22.56 & 42.93 \\ \hline
\end{tabular}
\caption{Betatron phases of Fuji- and D5-RFSoC relative to the midpoint between D10\_L02 and D10\_L03.}
\label{table:D10_L02phase}
\end{table}

Furthermore, as seen in Fig.~\ref{fig:first_oscillation}, horizontal oscillations tend to begin earlier than vertical oscillations.
To investigate the cause of this behavior, we also examine the possible influence of the dispersion function.
As shown in Fig.~\ref{fig:lattice_D10_L02}~(bottom), the horizontal dispersion at the midpoint between D10\_L02 and D10\_L03 is non-zero.
If the beam interacts with dust or other material at this location and loses momentum, horizontal oscillations can be excited due to the dispersion.
A similar phenomenon was observed during SuperKEKB operation in 2016, in which energy loss led to synchrotron oscillations~\cite{Terui:2018pae}.
By analogy, beam–dust interactions during SBL events could also cause momentum loss, resulting in horizontal oscillations.
The longitudinal momentum loss may enhance horizontal oscillations and could be responsible for their earlier appearance compared to vertical oscillations.

\subsection{Potential scenario of SBL evolution arising at D10\_L05}
\begin{figure}[hbt]
 \centering
 \includegraphics[width=0.8\columnwidth]{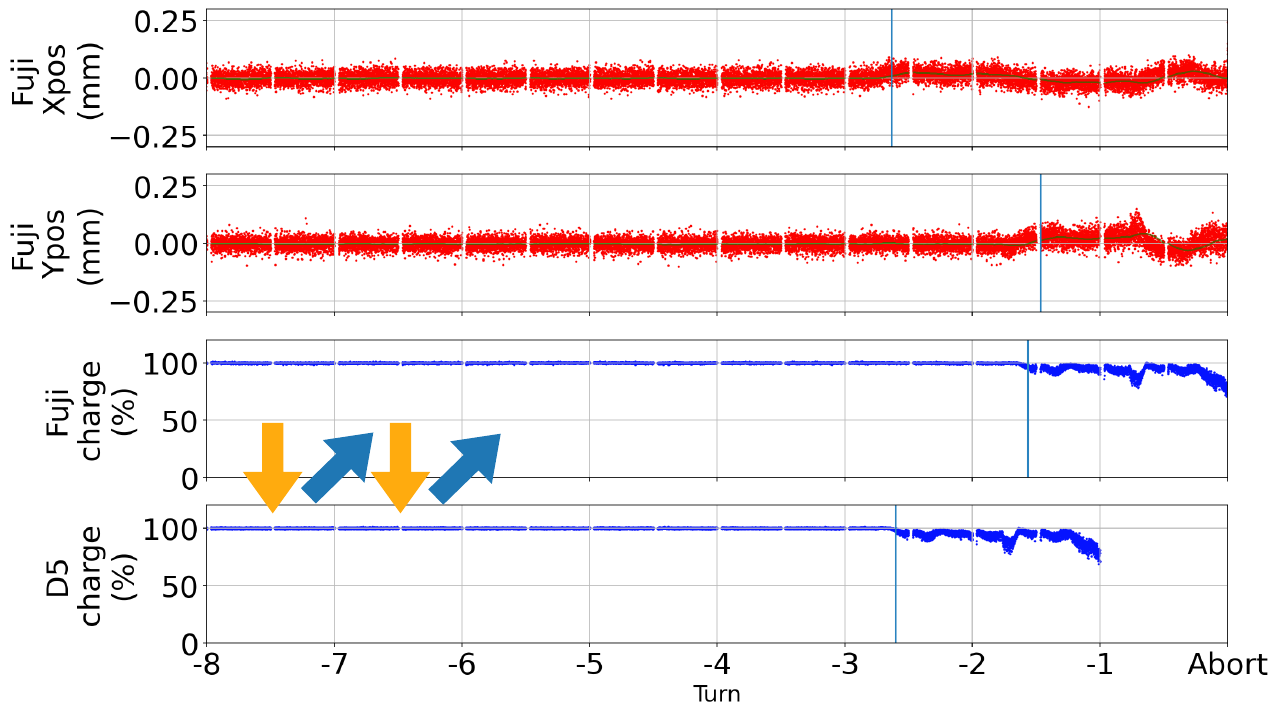}
 \caption{An SBL event accompanied by a pressure burst at D10\_L05. From top to bottom: Horizontal (X) position, vertical (Y) position, and bunch charge recorded by the Fuji-RFSoC, and bunch charge recorded by the D5-RFSoC. Vertical blue lines indicate the detected onset times of oscillation and charge loss.}
 \label{fig:sbl_10-29-194912}
\end{figure}

A characteristic feature of SBL events accompanied by pressure bursts at D10\_L05 is the short duration of bunch position oscillation before beam abort.  
Figure~\ref{fig:sbl_10-29-194912} shows an example of such an event.
This event occurred at 19:49:12 on October 29, 2024, with 2346 bunches circulating in the ring and an average bunch current of 0.51~mA. 
Since no position oscillations were observed at D5-RFSoC before significant charge loss, its position data are omitted.

As seen in Fig.~\ref{fig:sbl_10-29-194912}, the onset of horizontal oscillation at the Fuji-RFSoC occurs nearly simultaneously with the onset of charge loss at the D5-RFSoC.  
This means that the bunches whose oscillations were observed when they passed through the Fuji-RFSoC, have already lost their charge when they passed through Section 1 and arrived at the D5-RFSoC.
The charge was still intact when the oscillations were observed in Fuji-RFSoC, so the charge loss must have occurred in Section~1, specifically in the D06 collimator section.
The bunch appears to strike the collimators in D06 before completing a full turn around the ring.  
This explains the short delay between the onset of oscillation and the beam loss detected by the loss monitor, which in turn results in the short oscillation duration observed in Fig.~\ref{fig:duration_pressure}.

\begin{figure}[hbt]
 \centering
 \begin{subfigure}[b]{0.45\textwidth}
 \includegraphics[width=1\columnwidth]{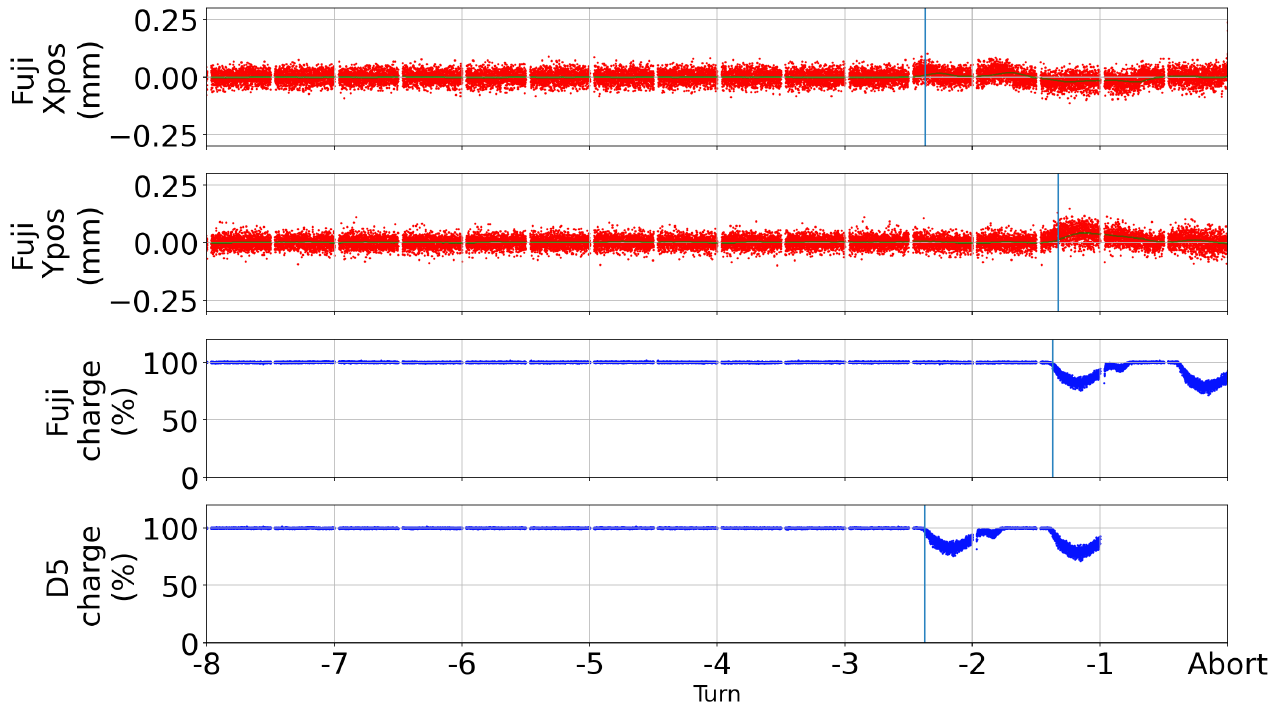}
 \subcaption{SBL observed at 21:28:33 on October 27, 2024. Num of bunches: 2346; average bunch current: 0.49~mA.}
 \label{fig:sbl_10-27-212833}
\end{subfigure}
\hfill
\begin{subfigure}[b]{0.45\textwidth}
 \centering
 \includegraphics[width=1\columnwidth]{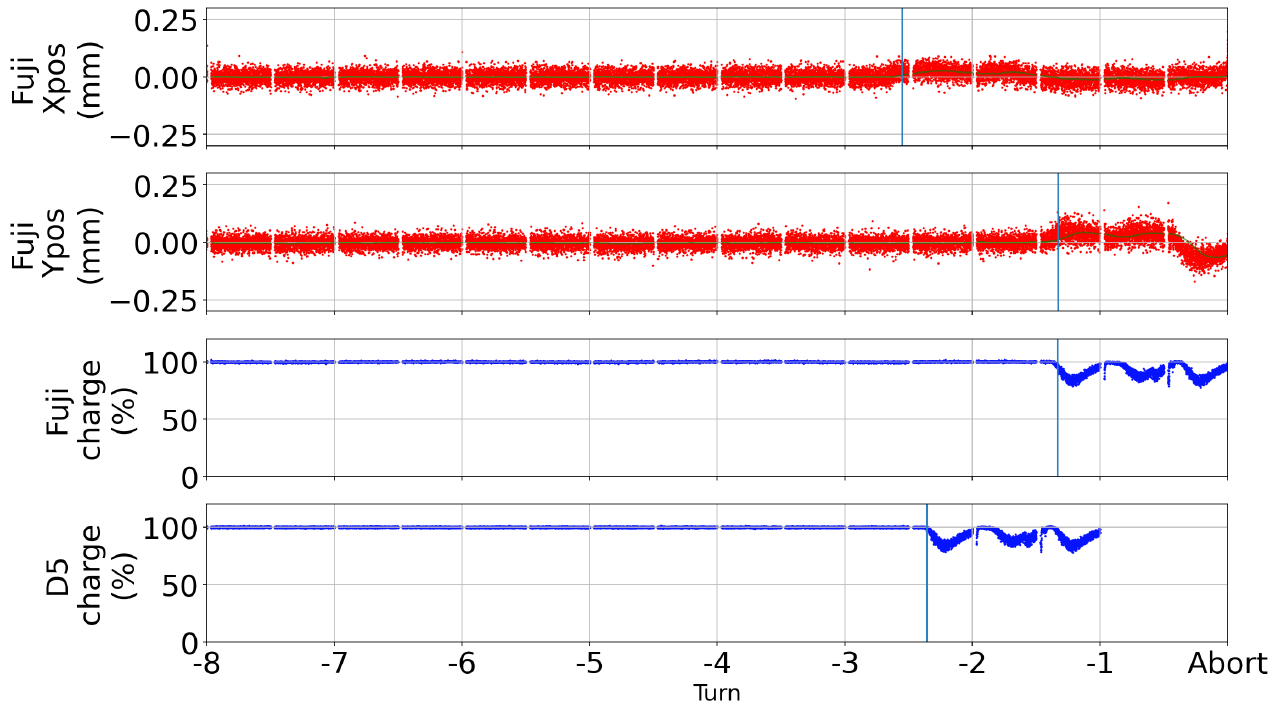}
 \subcaption{SBL observed at 02:30:07 on October 28, 2024. Num of bunches: 2346; average bunch current: 0.50~mA.}
 \label{fig:sbl_10-28-023007_2}
\end{subfigure}
\hfill
\begin{subfigure}[b]{0.45\textwidth}
 \centering
 \includegraphics[width=1\columnwidth]{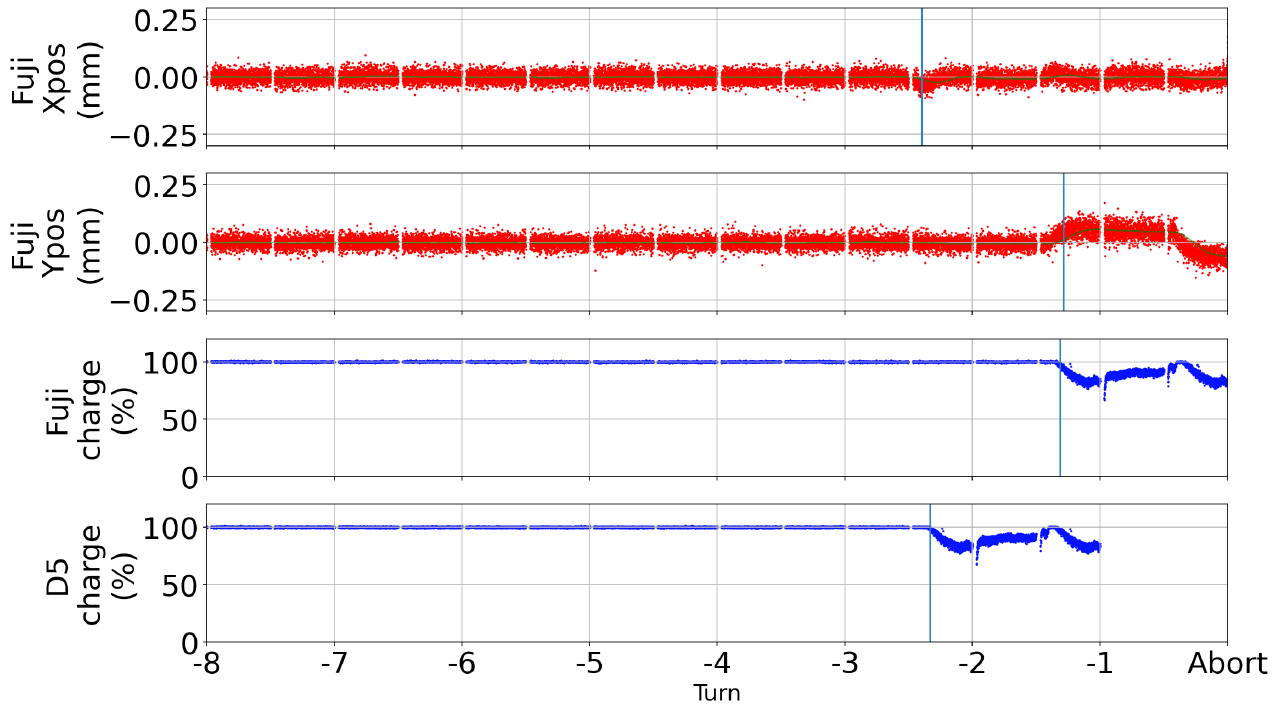}
 \subcaption{SBL observed at 02:30:10 on October 30, 2024. Num of bunches: 2346; average bunch current: 0.55~mA.}
 \label{fig:sbl_10-30-023010}
\end{subfigure}
\caption{SBL events accompanied by pressure bursts at D10\_L05.}
\label{fig:sbl_d10_l05}
\end{figure}

Figure~\ref{fig:sbl_d10_l05} presents three additional SBL events associated with pressure bursts at D10\_L05.  
The format is identical to Fig.~\ref{fig:sbl_10-29-194912}.  
In all cases, the relationship between the oscillation and charge loss onset timings at the Fuji- and D5-RFSoC is remarkably consistent.

\begin{figure}[hbt]
 \centering
 \includegraphics[width=0.8\columnwidth]{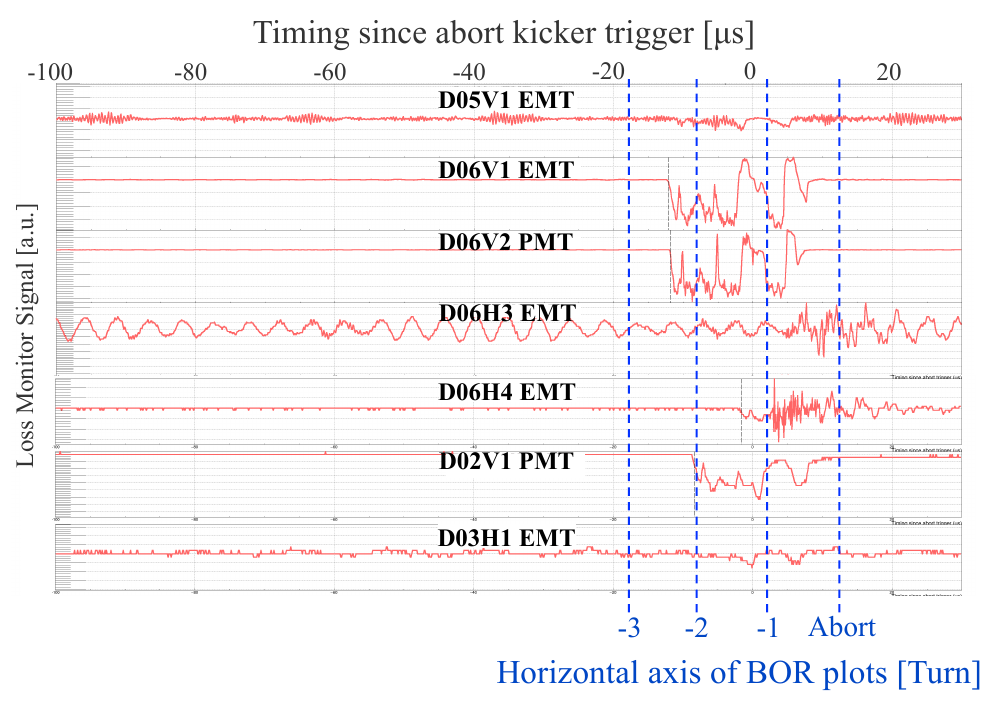}
 \caption{Loss monitor signals recorded during the SBL event at 19:49:12 on October 29, 2024. Each trace represents the signal from a PMT or EMT installed near a collimator. Radiation from beam loss increases signal amplitude. The time axis is defined such that \(t=0\) corresponds to the beam abort trigger issued by the central control room. Blue dashed lines indicate the same timing as in Fig.~\ref{fig:sbl_10-29-194912} for comparison.}
 \label{fig:lmtiming}
\end{figure}

To understand the sequence of events, we analyze which collimator in the D06 section the bunch hits first, based on fast loss monitor signals.  
In addition to the loss monitors for beam abort described in Section~\ref{lossmonitor_and_beamabortsystem}, several fast-response loss monitors dedicated to SBL observation are installed near collimators~\cite{Yoshihara:2024vme}.  
These include cesium iodide scintillator with photomultiplier tube (CsI+PMT) monitors at D02V1 and D06V2, and electron multiplier tube (EMT) monitors at D06H3, D06H4, D06V1, D05V1, and D03H1.
Figure~\ref{fig:lmtiming} shows the loss monitor signals recorded during the event in Fig.~\ref{fig:sbl_10-29-194912}~\cite{xiaodong,Yoshihara:2024vme}.  
The blue dashed vertical lines indicate the same timing as the turn axis in Fig.~\ref{fig:sbl_10-29-194912}.  
According to this figure, the loss monitor installed at D06V1 detected radiation earliest—approximately 2.5 turns before the beam abort trigger.
This closely follows the onset of horizontal oscillation observed at the Fuji-RFSoC.  
It thus appears that the bunch, already oscillating at the Fuji-RFSoC, first collided with the vertical collimator D06V1.  
Although D06H3 and D06H4 collimators (horizontal collimators located upstream of D06V1) also have loss monitors; they detected losses one turn later.  
This pattern was consistently observed in all three additional D10\_L05-related events.
These results suggest that vertical oscillations were likely initiated simultaneously with horizontal oscillations, and it was the vertical oscillation that first caused beam loss at D06V1.

However, examining the bunch position traces at the Fuji-RFSoC revealed that horizontal oscillations start first, while vertical oscillations appear one turn later.
Given that the pressure burst occurred at D10\_L05, it is plausible that the bunch experienced a kick at this location.  
Table~\ref{table:phase_D10_L05} shows the betatron phases at the Fuji-RFSoC and D06V1 relative to D10\_L05.
The vertical phase difference between D10\_L05 and the Fuji-RFSoC is nearly an integer multiple of \(\pi\), making vertical oscillation less visible at the Fuji-RFSoC.
This is consistent with the fact that vertical oscillations appeared to start later than horizontal oscillations at the Fuji-RFSoC.
Conversely, the vertical phase difference between D10\_L05 and D06V1 collimator is close to a half-integer multiple of \(\pi\), enhancing the visibility of oscillation at the collimator.  
This is consistent with the observation that the bunch first strikes the D06V1 collimator.

\begin{table}[h]
\centering
\begin{tabular}{c|ccc}
\hline
                     & D10\_L05 & Fuji-RFSoC & D06V1 Collimator \\ \hline
Horizontal betatron phase [rad/\(\pi\)] & 0        & 20.73     & 31.23 \\
Vertical betatron phase [rad/\(\pi\)] & 0        & 22.01     & 33.44 \\ \hline
\end{tabular}
\caption{Betatron phases of Fuji-RFSoC and D06V1 relative to D10\_L05.}
\label{table:phase_D10_L05}
\end{table}

\begin{figure}[hbt]
 \centering
 \includegraphics[width=0.6\columnwidth]{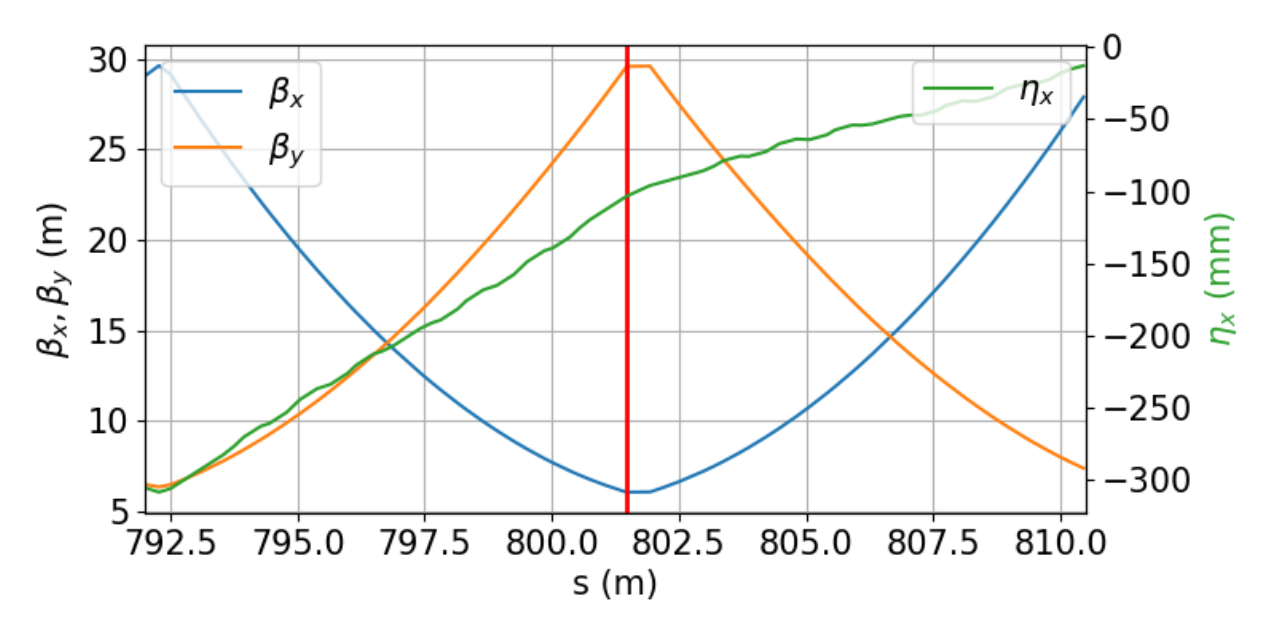}
 \caption{Beta and dispersion functions near D10\_L05. The horizontal axis indicates the distance from the collision point. The left vertical axis shows the beta function; the right shows the horizontal dispersion. The red vertical line marks the position of D10\_L05.}
 \label{fig:lattice_D10_L05}
\end{figure}

Figure~\ref{fig:lattice_D10_L05} shows the beta and dispersion functions near D10\_L05.  
A vertical focusing quadrupole magnet is located at D10\_L05, causing the vertical beta function to peak.  
According to Eq.~(\ref{eq:kick}), the larger the beta function at the kick location, the greater the resulting oscillation.  
Thus, a kick at D10\_L05 would significantly enhance vertical oscillation, potentially leading to immediate impact with vertical collimators and charge loss.

\section{Summary}
\label{04_Summary}

This study has significantly advanced our understanding of SBL and provided crucial insights for future SBL mitigation strategies and luminosity improvement efforts.

Specifically, observations using the BOR revealed that bunch charge loss during SBL events predominantly occurs at the collimators in the D06 section.  
This finding suggests that substantial beam position oscillations likely occur before the beam enters the D06 section.
Moreover, classifying SBL events based on the location of associated pressure bursts uncovered distinct patterns in the onset of oscillations.  
Further analysis from the viewpoint of the accelerator lattice implies a causal relationship wherein the beam receives a disturbance at the pressure burst location, subsequently leading to the development of SBL.
These results clearly demonstrate the effectiveness of the BOR as a diagnostic instrument and establish a foundation for future studies of SBL phenomena in nanobeam accelerators.

Going forward, a key challenge will be to generalize the interpretation of SBL evolution by applying the scenario investigated at each pressure burst site to other locations around the ring.  
By increasing the number and resolution of BORs, it will be possible to determine whether SBL is a location-independent phenomenon governed by universal parameters such as the beta or dispersion function, or whether it is strongly influenced by site-specific conditions not captured by lattice optics alone.
Additionally, we will also focus on the possible beam size growth, as suggested in Appendix~\ref{appendixA}, to deepen our understanding of the mechanism of SBL events.

In addition, compared with similar fast beam-loss phenomena at the LHC, known as UFOs, SBL at SuperKEKB proceeds on an even shorter timescale.
At the LHC, machine damage was mitigated by optimizing the beam loss monitor system~\cite{Auchmann:2016upc}; however, for SBL such mitigation is insufficient, as a beam abort based solely on loss monitors cannot be executed quickly enough.
This highlights the necessity of diagnostics with higher time resolution.
In this work, by focusing on bunch-by-bunch beam position measurements, we were able to directly observe the onset of bunch oscillations.
Furthermore, by combining these observations with pressure data, we found that the bunch oscillations are closely linked to pressure bursts.
These findings represent entirely new insights and may provide valuable observables not only for SuperKEKB but also for LHC and future facilities such as proposed Higgs factories.
Nevertheless, whether SBL is a dust-induced beam instability, analogous to UFOs at the LHC, cannot be determined from the present results alone.
To address this question, we plan to combine bunch position, pressure, and future beam size measurements with detailed simulations in order to move closer to the underlying mechanism of SBL.

\section{Acknowledgements}
R.~Nomaru and G.~Mitsuka work was supported by Japan Society for the Promotion of Science (JSPS) International Leading Research Grant Number JP22K21347. In addition, R.~Nomaru work was supported by JSPS Core-to-Core Program (Grant Number: JPJSCCA20230004). L.~Ruckman work was supported by the U.S. Department of Energy, under contract number DE-AC02-76SF00515.

\appendix
\section{Discussion on the amplitude of bunch position oscillations}
\label{appendixA}
In Section~\ref{Oscillation_amplitude}, we discussed the amplitude of bunch position oscillations during SBL events.  
Here, we further investigate this topic by comparing the observed oscillation amplitudes with the results of a previous test.

\begin{figure}[hbt]
 \centering
 \includegraphics[width=0.8\columnwidth]{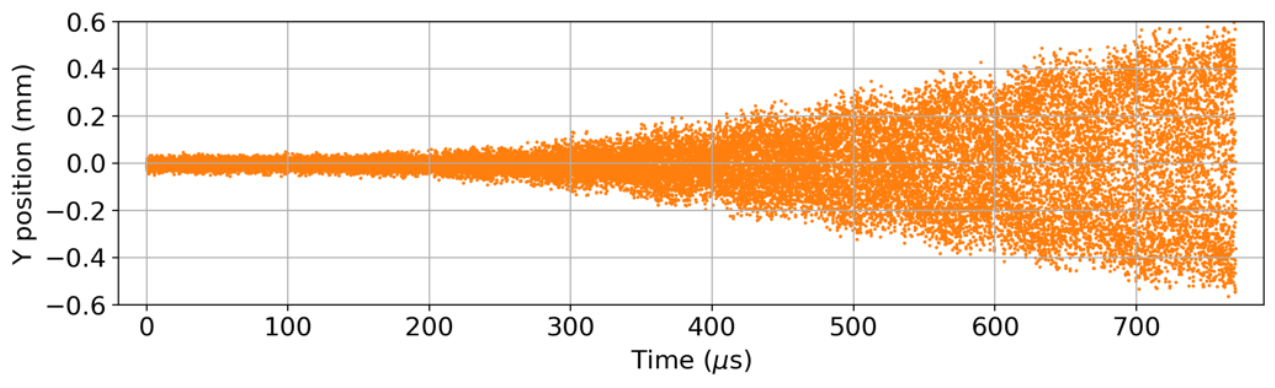}
 \caption{Results from the feedback system test. The horizontal axis represents elapsed time, and the vertical axis shows vertical bunch positions measured by the Fuji-RFSoC. Each dot corresponds to a single bunch.}
 \label{fig:feedback_test}
\end{figure}

During the development of the BOR, we conducted a performance test using the main ring's bunch-by-bunch feedback system.  
Details of this test are reported in Ref.~\cite{Nomaru:2024qls}.  
In this test, we reversed the phase of the feedback kicker in the bunch-by-bunch feedback system~\cite{Tobiyama:2016pasj} to deliberately amplify the bunch oscillations in a total of 393 bunches, and monitored the resulting motion using the BOR.  
The feedback kicker was installed in the Fuji straight section.  
The results of this test are shown in Fig.~\ref{fig:feedback_test}.
This figure demonstrates that BOR successfully captured the increase in bunch oscillation amplitude caused by the inverted kicker phase.  
A beam abort occurred after the right edge of the plot due to the excessive oscillation amplitude.
The optics used in this test was the $\beta_y^{\ast} = 1~\mathrm{mm}$ configuration, which features a relatively tight physical aperture.  
The strength of the kick applied in the test was equivalent to that applied under normal operating conditions, since the feedback gain settings remained unchanged.

According to Fig.~\ref{fig:feedback_test}, the bunch oscillations grew to approximately $\pm$0.6~mm (i.e., oscillation amplitude $\sim$1.2~mm).  
During the 750~µs interval over which the oscillation amplitude increased, no beam loss occurred, and measurement was uninterrupted.  
This indicates that, even under tight aperture optics with $\beta_y^{\ast} = 1~\mathrm{mm}$, oscillations up to roughly 1.2~mm are tolerable.
However, as shown in Fig.~\ref{fig:oscillation_amp}, oscillation amplitudes observed during SBL events are generally smaller than this.  
This implies that, despite the oscillation amplitude being within the physically allowed aperture, substantial charge loss and beam aborts frequently occur during SBL events.
This discrepancy suggests that beam loss during SBL may not be due solely to increased bunch position oscillations.  
Rather, it may result from a combination of position oscillation and simultaneous beam size growth.  
Indeed, beam size monitors have recorded signs of increased beam size during several SBL events~\cite{mitsuka:ipac2025-thps089}, supporting this interpretation.

In Fig.~\ref{fig:feedback_test}, a clear coherent growth of bunch position oscillations is observed, showing a pattern in which the oscillation appears to diffuse along the bunch train. In contrast, in several SBL events presented in this paper, the oscillation pattern along the bunch train is significantly different: the oscillations often appear as if the motion propagates along successive bunches. One possible explanation is that the feedback kicker applies kicks to each bunch in varying directions, whereas the kick source responsible for SBL may act over a certain period of time, continuously affecting multiple bunches. These observations suggest that the feedback system, including the feedback kicker, is unlikely to be the origin of SBL.

\bibliographystyle{unsrt}
\bibliography{bibliography}

@PREAMBLE{
 "\providecommand{\noopsort}[1]{}" 
 # "\providecommand{\singleletter}[1]{#1}%" 
}

@misc{SuperKEKBTDR,
    title="{SuperKEKB Design Report}",
    howpublished = {\url{https://kds.kek.jp/event/15914/}},
}

@article{Nomaru:2024qls,
    author = "Nomaru, R. and Mitsuka, G. and Ruckman, L. and Herbst, R.",
    title = "{Development of a novel bunch oscillation recorder with RFSoC technology}",
    doi = "10.1088/1748-0221/19/12/P12026",
    journal = "JINST",
    volume = "19",
    number = "12",
    pages = "P12026",
    year = "2024"
}

@article{Ishibashi:2020rgt,
    author = "Ishibashi, T. and Terui, S. and Suetsugu, Y. and Watanabe, K. and Shirai, M.",
    title = "{Movable collimator system for SuperKEKB}",
    doi = "10.1103/PhysRevAccelBeams.23.053501",
    journal = "Phys. Rev. Accel. Beams",
    volume = "23",
    number = "5",
    pages = "053501",
    year = "2020"
}

@inproceedings{mimashi:ipac14-mopro023,
    author = {T. Mimashi and others},
    title = {{SuperKEKB Beam Abort System}},
    booktitle = {Proc. IPAC'14},
    pages = {116--118},
    paper = {MOPRO023},
    doi = {10.18429/JACoW-IPAC2014-MOPRO023},
    url = {https://jacow.org/IPAC2014/papers/MOPRO023.pdf},
    year="2014"
}

@inproceedings{Terui:2018pae,
    author = "Terui, S. and Hisamatsu, H. and Ishibashi, T. and Kanazawa, K. and Shibata, K. and Shirai, M. and Suetsugu, Y.",
    title = "{Observation of Pressure Bursts in the SuperKEKB Positron Ring}",
    booktitle = "{9th International Particle Accelerator Conference}",
    doi = "10.18429/JACoW-IPAC2018-WEPML058",
    month = "6",
    year = "2018"
}

@article{Yoshihara:2024vme,
    author = "Yoshihara, K. and others",
    title = "{Development and implementation of advanced beam diagnostic and abort systems in SuperKEKB}",
    doi = "10.1016/j.nima.2024.170117",
    journal = "Nucl. Instrum. Meth. A",
    volume = "1072",
    pages = "170117",
    year = "2025"
}

@misc{xiaodong,
  author       = {X. Shi},
  title        = {{Private communication}},
  year         = {2024},
}

@inproceedings{ikeda:ibic14-tupd22,
    author = {H. Ikeda and M. Arinaga and J. W. Flanagan and H. Fukuma and M. Tobiyama},
    title = {{Beam Loss Monitor at SuperKEKB}},
    booktitle = {Proc. IBIC'14},
    pages = {459--462},
    paper = {TUPD22},
    url = {https://jacow.org/IBIC2014/papers/TUPD22.pdf},
    year = "2014"
}

@article{Bacher:2021frk,
    author = "Bacher, S. and others",
    title = "{Performance of the diamond-based beam-loss monitor system of Belle II}",
    doi = "10.1016/j.nima.2021.165157",
    journal = "Nucl. Instrum. Meth. A",
    volume = "997",
    pages = "165157",
    year = "2021"
}

@INPROCEEDINGS{claws_proc,
  author={Gabriel, M. and Kattau, M. and Kiesling, C. and Simon, F. and Windel, H.},
  booktitle={2016 IEEE Nuclear Science Symposium, Medical Imaging Conference and Room-Temperature Semiconductor Detector Workshop (NSS/MIC/RTSD)}, 
  title={{CLAWS — a plastic scintillator / SiPM based detector to measure backgrounds at SuperKEKB}}, 
  year={2016},
  volume={},
  number={},
  pages={1-6},
  doi={10.1109/NSSMIC.2016.8069685}}

@article{Terui:2024jue,
    author = "Terui, S. and others",
    title = "{Construction of SuperKEKB vacuum control system and its eight years operation experience}",
    doi = "10.1016/j.nima.2024.169814",
    journal = "Nucl. Instrum. Meth. A",
    volume = "1068",
    pages = "169814",
    year = "2024"
}

@inproceedings{Tobiyama:2016pasj,
    author = "M. Tobiyama and J.W. Flanagan",
    title = "{Bunch by bunch feedback systems for SuperKEKB rings}",
    booktitle = "{Proceedings of the 13th Annual Meeting of Particle Accelerator Society of Japan}",
    year = "2016"
}

@inproceedings{mitsuka:ipac2025-thps089,
    author = {G. Mitsuka and others},
    title = {{Measurements for beam size blowup in sudden beam loss events and analysis of the beam loss evolution mechanism}},
    booktitle = {Proc. IPAC'25},
    pages = {3144-3147},
    paper = {THPS089},
    year = {2025},
    doi = {10.18429/JACoW-IPAC25-THPS089},
}

@inproceedings{ikeda:ipac2025-mocd3,
    author = {H. Ikeda and others},
    title = {{Observations and efforts to reduce sudden beam loss at SuperKEKB}},
    booktitle = {Proc. IPAC'25},
    pages = {57-60},
    paper = {MOCD3},
    year = {2025},
    doi = {10.18429/JACoW-IPAC25-MOCD3},
}

@article{Suetsugu:2003gb,
    author = "Suetsugu, Y. and Kageyama, T. and Shibata, K. and Sanami, T.",
    title = "{Latest movable mask system for KEKB}",
    doi = "10.1016/j.nima.2003.06.003",
    journal = "Nucl. Instrum. Meth. A",
    volume = "513",
    pages = "465--472",
    year = "2003"
}

@article{10.1093/ptep/ptz106,
    author = {Kou, E and others},
    title = {{The Belle II Physics Book}},
    journal = {Progress of Theoretical and Experimental Physics},
    volume = {2019},
    number = {12},
    pages = {123C01},
    year = {2019},
    month = {12},
    issn = {2050-3911},
    doi = {10.1093/ptep/ptz106},
    url = {https://doi.org/10.1093/ptep/ptz106},
    eprint = {https://academic.oup.com/ptep/article-pdf/2019/12/123C01/32693980/ptz106.pdf},
}

@article{Lindstrom:2020hks,
    author = "Lindstrom, B. and others",
    title = "{Dynamics of the interaction of dust particles with the LHC beam}",
    doi = "10.1103/PhysRevAccelBeams.23.124501",
    journal = "Phys. Rev. Accel. Beams",
    volume = "23",
    number = "12",
    pages = "124501",
    year = "2020"
}

@article{Baer:2011mf,
    author = "Baer, T. and others",
    editor = "Petit-Jean-Genaz, Christine",
    title = "{UFOs in the LHC}",
    reportNumber = "IPAC-2011-TUPC137",
    journal = "Conf. Proc. C",
    volume = "110904",
    pages = "1347--1349",
    year = "2011"
}

@inproceedings{nomaru:ipac2025-thpm089,
    author = {Nomaru, R. and Mitsuka, G. and Ruckman, L.},
    title = "{Disentangling sudden beam loss events and fast beam abort system with the RFSoC-BPM at SuperKEKB}",
    booktitle = {Proc. IPAC'25},
    pages = {2870-2873},
    paper = {THPM089},
    year = {2025},
    doi = {10.18429/JACoW-IPAC25-THPM089}
}

@inproceedings{Auchmann:2016upc,
    author = "Auchmann, B. and Ghini, J. and Grob, L. and Iadarola, G. and Lechner, A. and Papotti, G.",
    title = "{How to survive a UFO attack}",
    booktitle = "{6th Evian Workshop on LHC beam operation}",
    publisher = "CERN",
    address = "Geneva",
    pages = "81--86",
    year = "2016"
}

@article{Mulyani:2019gsy,
    author = "Mulyani, E. and Flanagan, J.W. and Tobiyama, M. and Fukuma, H. and Ikeda, H. and Mitsuka, G.",
    title = "{First measurements of the vertical beam size with an X-ray beam size monitor in SuperKEKB rings}",
    doi = "10.1016/j.nima.2018.11.116",
    journal = "Nucl. Instrum. Meth. A",
    volume = "919",
    pages = "1--15",
    year = "2019"
}

@misc{SuperKEKB_status_and_plan,
  title        = "{SuperKEKB Operation Status and Plan}",
  howpublished = {\url{https://www-linac.kek.jp/skekb/status/web/status_plan.md.html}},
}

@misc{Ohnishi:eeFACT25,
  author       = {Y. Ohnishi},
  title        = "{Recent Performance of SuperKEKB}",
  howpublished = {Presentation at eeFACT'25},
  year         = {2025},
}

\end{document}